\documentclass[conference]{IEEEtran}
\IEEEoverridecommandlockouts
\pdfoutput=1
\usepackage{graphicx}
\usepackage{amsmath}
\usepackage{amssymb}
\usepackage[noadjust]{cite}
\usepackage{float}
\usepackage{color}
\usepackage{url}
\usepackage{dblfloatfix} 
\usepackage{xfrac}
\usepackage{amsfonts}
\usepackage{enumitem}
\usepackage{float}
\usepackage{color}
\usepackage{cite}
\usepackage{graphicx}
\usepackage{epstopdf}
\usepackage{bm}
\usepackage{subcaption}
\usepackage{relsize}
\usepackage{mathtools}
\usepackage{array}
\usepackage{array}
\usepackage{multirow}
\usepackage{multicol}
\usepackage{mathtools}

\usepackage{blindtext}
\usepackage{caption}
  \usepackage{comment}
  \usepackage{lipsum}
\usepackage{xcolor}
\usepackage{amssymb}
\usepackage{pifont}
\newcolumntype{P}[1]{>{\centering\arraybackslash}p{#1}}
\everypar{\looseness=-1}
\usepackage{soul}
\graphicspath{{Figures/}}

\begin{document}
\title{Indoor Propagation Measurements with {Sekisui} Transparent Reflectors at 28/39/120/144 GHz}
\author{Chethan K. Anjinappa, 
Ashwini P. Ganesh, Ozgur Ozdemir, Kris Ridenour, Wahab Khawaja, \.{I}smail G\"{u}ven\c{c},\\ Hiroyuki Nomoto$^\dagger$, and Yasuaki Ide$^\dagger$\\
Department of Electrical and Computer Engineering, NC State University, Raleigh, NC 27606\\
$^\dagger$Research \& Development Institute, High Performance Plastics Company, Sekisui Chemical CO.,LTD.\\
e-mail:~\{canjina, apondey, oozdemi, kwrideno, wkhawaja, iguvenc\}@ncsu.edu \\
$^\dagger$\{hiroyuki.nomoto, yasuaki.ide\}@sekisui.com
\thanks{This work has been supported in part by the National Science Foundation (NSF) through the award  CNS-1916766 as well as the industry membership fees from the Broadband Wireless Access Center (BWAC) I/UCRC Center.} 
}

\renewcommand{\baselinestretch}{.93}

\maketitle
\begin{abstract}
One of the critical challenges of operating with the terahertz or millimeter-wave wireless networks is the necessity of at least a strong non-line-of-sight (NLoS) reflected path to form a stable link. Recent studies have shown that an economical way of enhancing/improving these NLoS links is by using passive metallic reflectors that provide strong reflections. However, despite its inherent radio advantage, metals can dramatically influence the landscape's appearance - especially the indoor environment. A conceptual view of escaping this is by using \textit{transparent reflectors}. In this work, for the very first time, we evaluate the wireless propagation characteristics of passive transparent reflectors in an indoor environment at~28~GHz,~39~GHz,~120~GHz, and~144~GHz bands. In particular, we investigate the penetration loss and the reflection characteristics at different frequencies and compare them against the other common indoor materials such as ceiling tile, clear glass, drywall, plywood, and metal. The measurement results suggest that the transparent reflector, apart from an obvious advantage of transparency, has a higher penetration loss than the common indoor materials (excluding metal) and performs similarly to metal in terms of reflection. Our experimental results directly translate to better reflection performance and preserving the radio waves within the environment than common indoor materials, with potential applications in controlled wireless communication.
\end{abstract}

\begin{IEEEkeywords}
~6G, B5G, Indoor coverage, mmWave, non-line-of-sight (NLOS), Sub-THz, Transparent reflectors.
\end{IEEEkeywords}

\section{Introduction}\label{Sec:Intro}
\IEEEPARstart{M}{\lowercase{illimeter wave}} (mmWave) and Terahertz bands are crucial enablers of beyond 5G (B5G)/6G wireless communications. In particular, the~THz communication is recognized as one of the nine key communication technology enablers~\cite{ComSoC_9KeyEnablers_6G}, due to its large bandwidth and its support of new possibilities ranging from augmented/virtual reality to high-resolution positioning and autonomous driving~\cite{chaccour2021seven}. One of the numerous challenges in a THz/mmWave network, is the requirement of the line of sight (LoS) or at least a strong first-order reflected non-LoS (NLoS) path to form a link, as diffraction is not significant at these frequencies. Particularly the availability of the NLoS link depends on the reflection profile of a scatterer, material, and the frequency of operation. 

There are numerous solutions proposed in the literature for mmWave indoor and outdoor NLoS coverage enhancement. These include utilizing high sensitivity receivers, high transmit power, use of multiple access points, and beam-forming using antenna arrays. However, all these solutions have limitations when there are blockage and NLoS propagation~\cite{khawaja2020coverage}. Re-configurable reflecting surfaces comprising of meta-materials and active repeaters for mmWave coverage enhancement have recently attracted extensive interest in the literature~\cite{RIS, ozdemir202028}. However, maintaining a large number of active repeaters and re-configurable reflecting surfaces is expensive, and the acquisition of channel state information is a complex problem~\cite{IRS_CSI}.

A simple and economical solution for mmWave coverage enhancement at indoor and outdoor scenarios is by using passive metallic reflectors~\cite{urban_reflector,khawaja2020coverage}. The metallic reflectors of different shapes and sizes can be selected based on the coverage requirement. For example, in \cite{khawaja2018coverage}, a flat metallic sheet helps to achieve directional coverage at 28~GHz deep in the corridor, whereas cylindrical and spherical reflectors help in a uniform but short-area corridor coverage. The different shapes and size reflectors can be made part of the interior/exterior design. Earlier simulation and measurement studies, such as~\cite{khawaja2020coverage,urban_reflector,Ender2020_Raytracing_RWS}, have shown that objects such as metal act as perfect reflectors, enabling strong reflections for directional NLoS communication. Thus, the deployment of the passive metallic reflectors can create a favorable propagation environment by introducing new multipath components (MPCs) to the channel and increasing the overall spatial diversity of the MPCs. However, metals will have a dramatic influence on appearance, especially the indoor environment, spoiling the landscape. For instance, attaching metals to windows blocks sunlight and view; one can think of many other possibilities. 

A viable option to counter the above disadvantage is the use of \textit{transparent reflectors}. These passive reflectors are not only transparent by design but are also thin, flexible, low-cost, and lightweight~\cite{Sekisui_Webpage}. Thus, they offer a more sustainable and easy-to-deploy alternative without altering the appearance of the surface. Even though this is fascinating from a conceptual view, how effective the transparent reflectors are in terms of radio propagation needs to be evaluated from a practical point of view. With these in perspective, the main goal of this paper is to investigate the critical wireless propagation characteristics of transparent reflectors. In particular, {we characterize the penetration loss (attenuation through the material), a measure signifying the leakiness of a material,} and reflection properties of the transparent reflector and evaluate its efficacy against the commonly used indoor material~\cite{kairui}. In terms of the operational frequency, we are primarily interested in the commonly considered mmWave bands of 28~GHz and 39~GHz, and the popular sub-THz bands of 120~GHz and 144~GHz. To the best of our knowledge, this is the first work to evaluate transparent reflectors for wireless communication at mmWave and sub-THz bands. Experimental results demonstrate the promising performance of the transparent reflectors compared to commonly found indoor materials with potential applications to diverse scenarios in the 5G and 6G wireless communications, such as indoor and outdoor coverage improvement and enhancement.

\section{System Model}\label{Sec:Sys_Model}
This section describes the details of the transparent reflectors and other indoor materials used in our measurements. We also briefly overview the channel sounder and its capabilities, and pivotal details for better understanding the penetration loss and reflection experiments.

\subsection{Transparent Reflectors}\label{Sec:Transparent_Reflector}
\begin{figure}[t!]
\centering
  {\includegraphics[width=.23\textwidth,height=.15\textwidth]{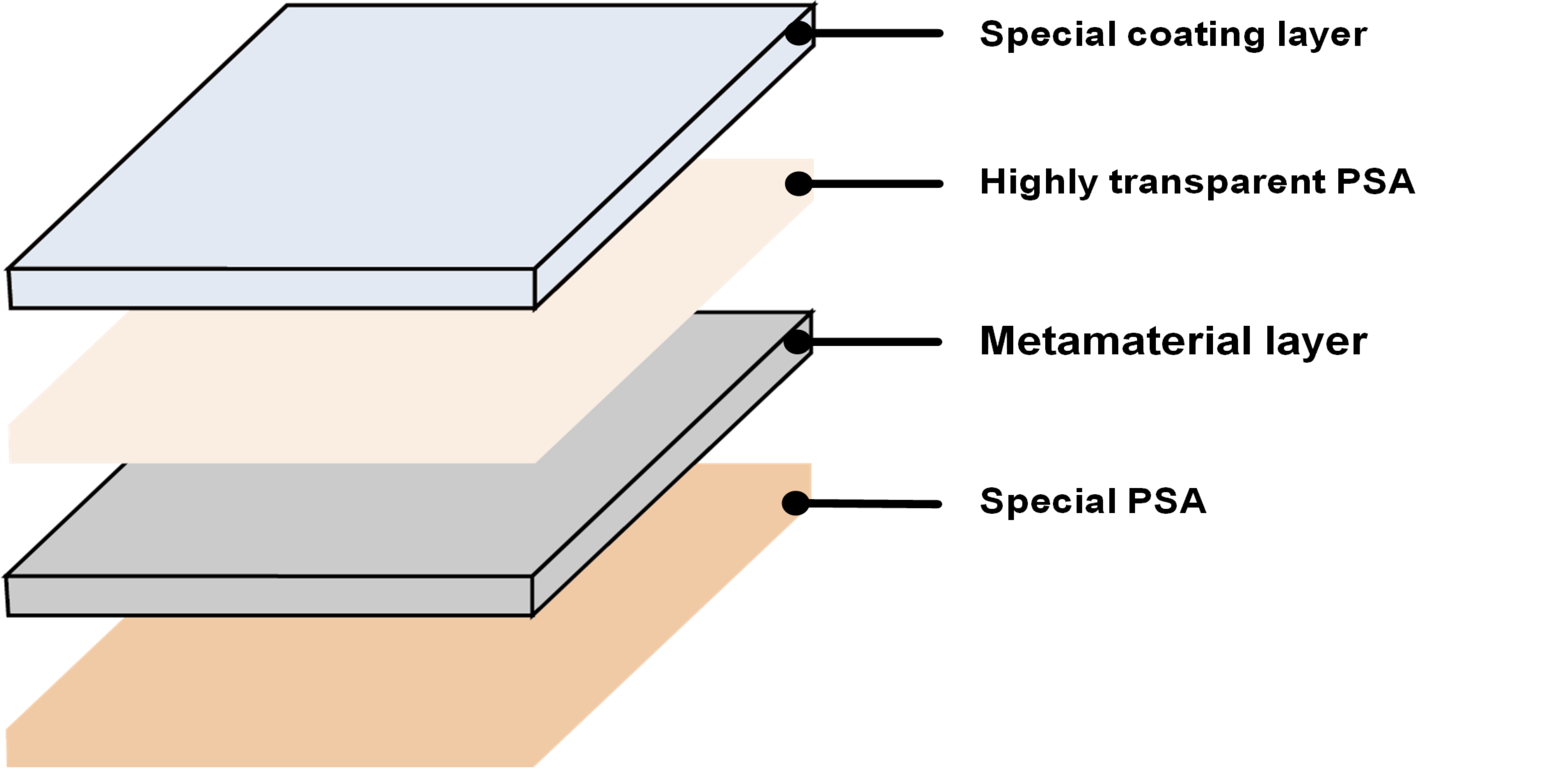}}\quad
{\includegraphics[width=.23\textwidth,height=.175\textwidth]{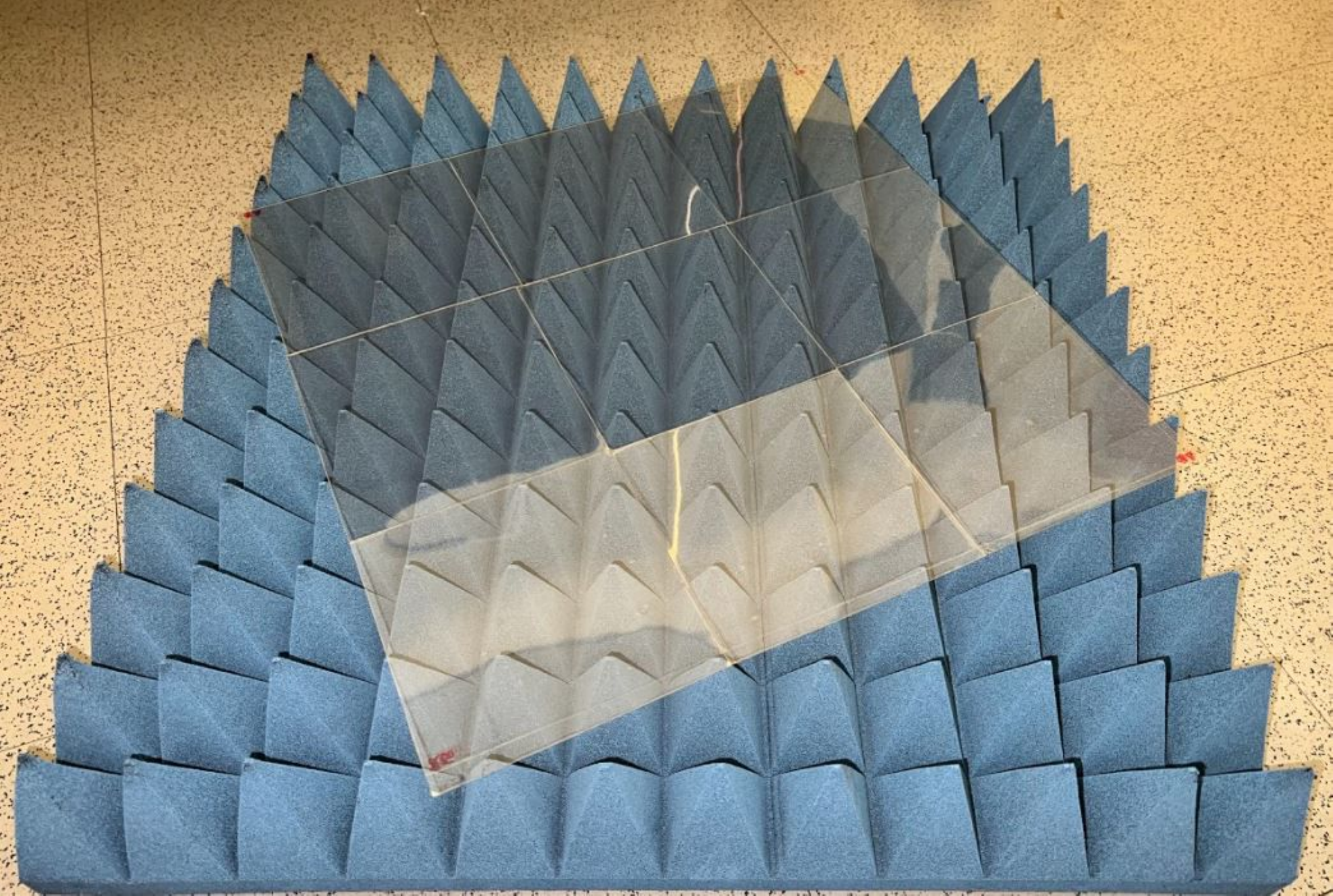}}\quad
\caption{ Transparent reflector film construction and appearance.}
  \label{fig:ref_details}  
\end{figure}

Transparent reflectors are flexible radio wave reflection films and are an economical and more sustainable alternative to conventional methods of increasing the signal strength. In terms of the construction, a transparent film consists of a meta-material layer that is structured to diffusely reflects high-frequency radio waves, a highly transparent adhesive, a special coating to protect the film's surface, and an adhesive layer at the bottom of the surface. This is illustrated graphically in Fig.~\ref{fig:ref_details}~\cite{Sekisui_Webpage}. These reflectors when placed on any surface such as a wall or ceiling will reflect all the incident rays enabling them to enhance and improve coverage. In this paper, we use transparent reflectors manufactured by Sekisui Chemical, Inc.

\subsection{Common Indoor Materials}\label{Sec:Common_Materials}
\begin{figure}[!t]
\centering
   \subfloat[Ceiling tile.]{\includegraphics[width=.23\textwidth,height=.18\textwidth]{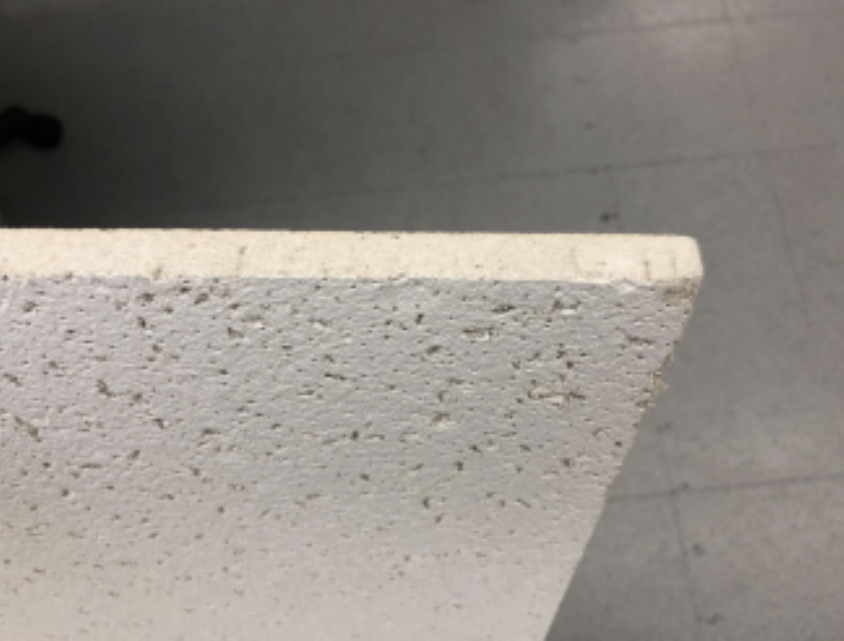}}\quad
   \subfloat[Clear glass.]{\includegraphics[width=.23\textwidth,height=.18\textwidth]{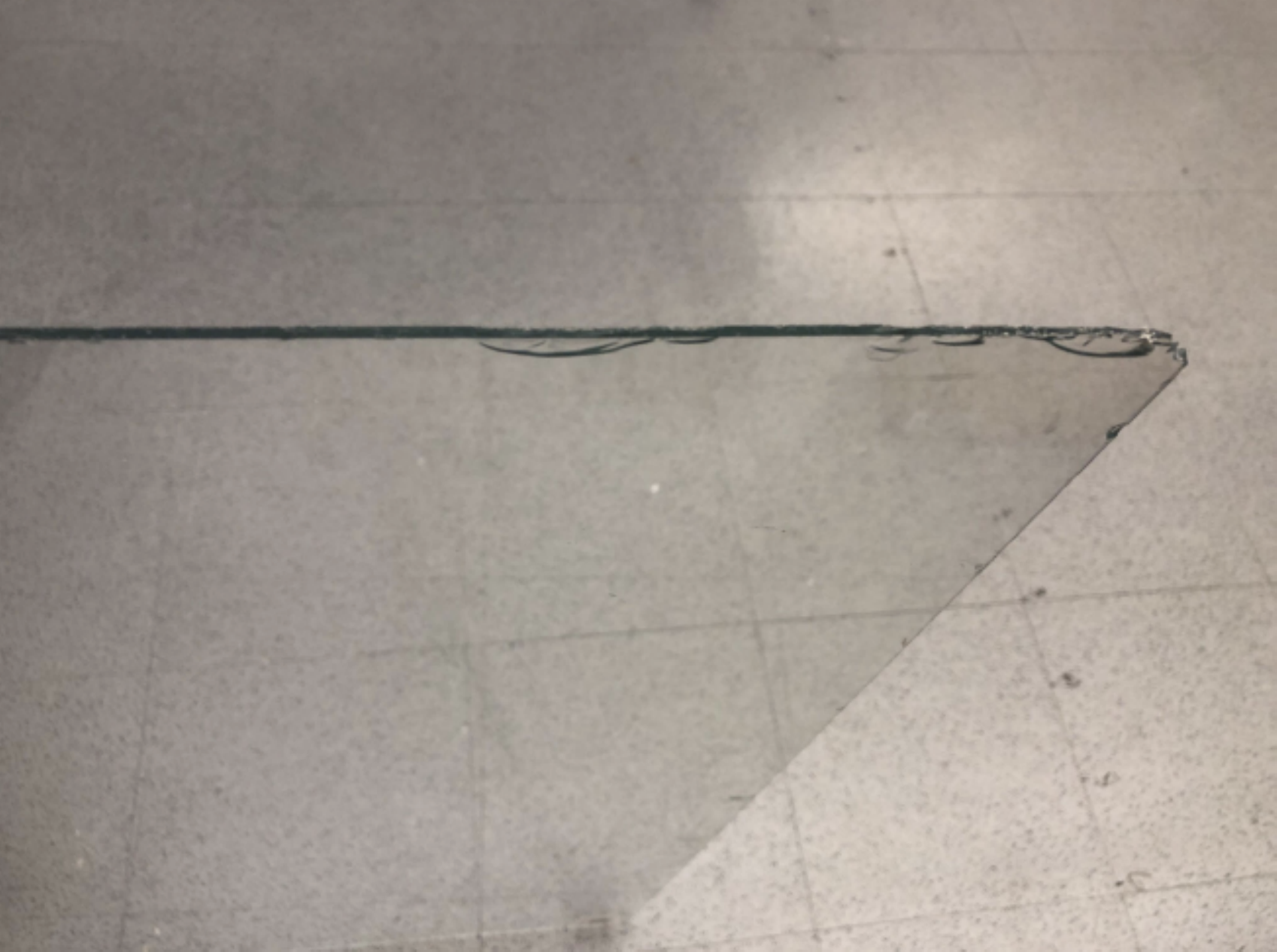}}\\
   \subfloat[Drywall.]{\includegraphics[width=.23\textwidth,height=.18\textwidth]{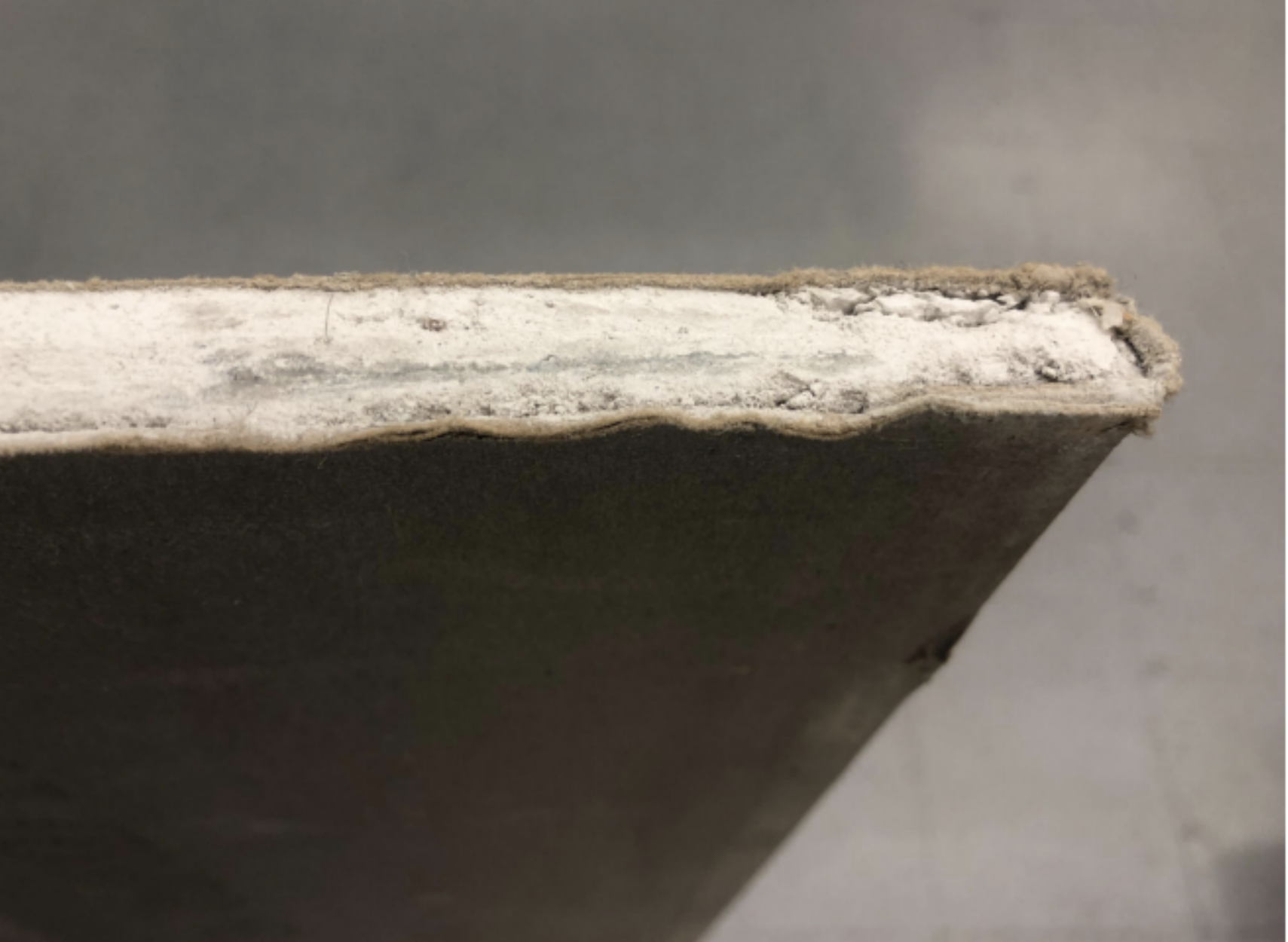}}\quad
   \subfloat[Plywood.]{\includegraphics[width=.23\textwidth,height=.18\textwidth]{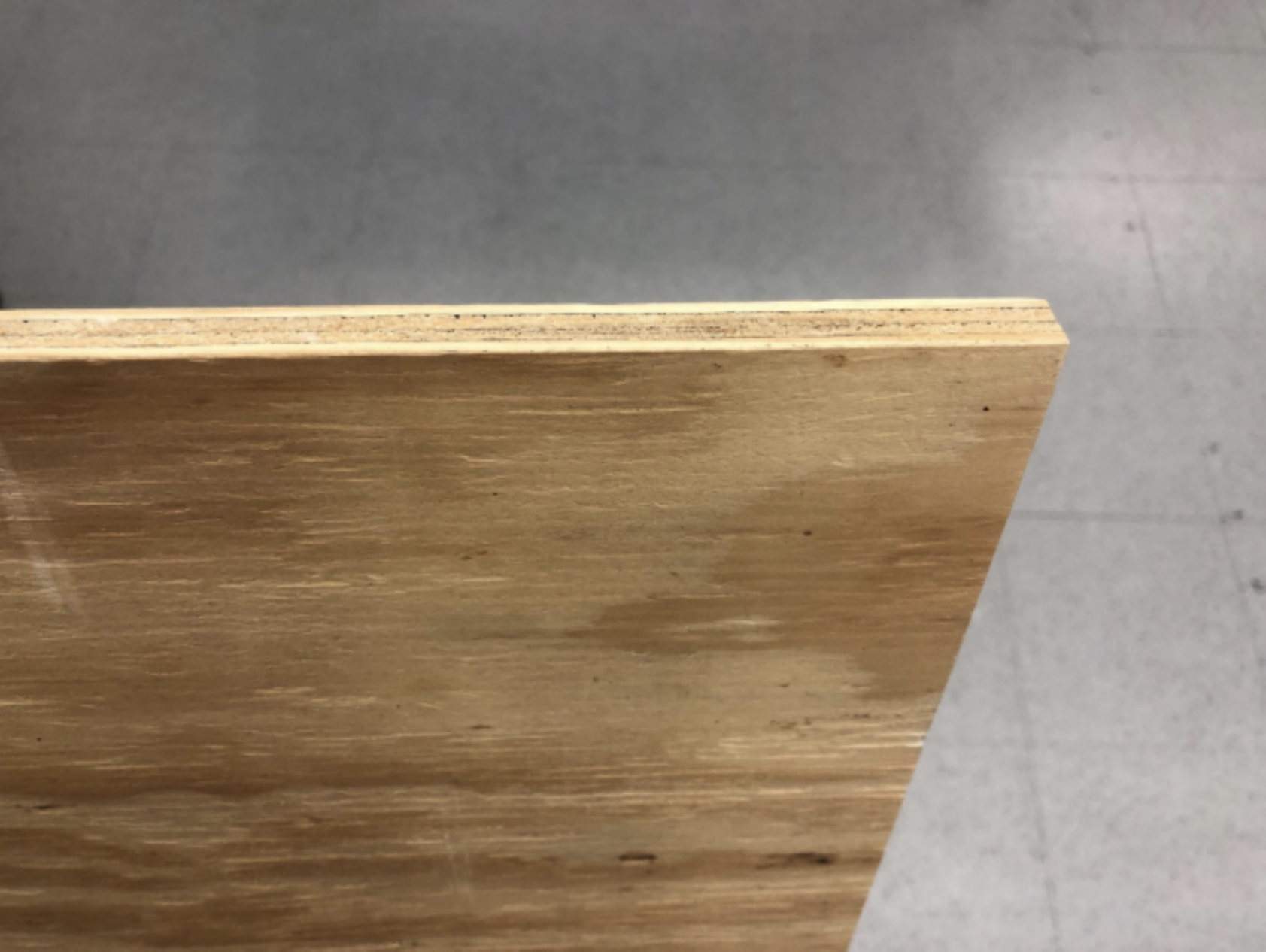}}\\
   \subfloat[Metal.]{\includegraphics[width=.23\textwidth,height=.18\textwidth]{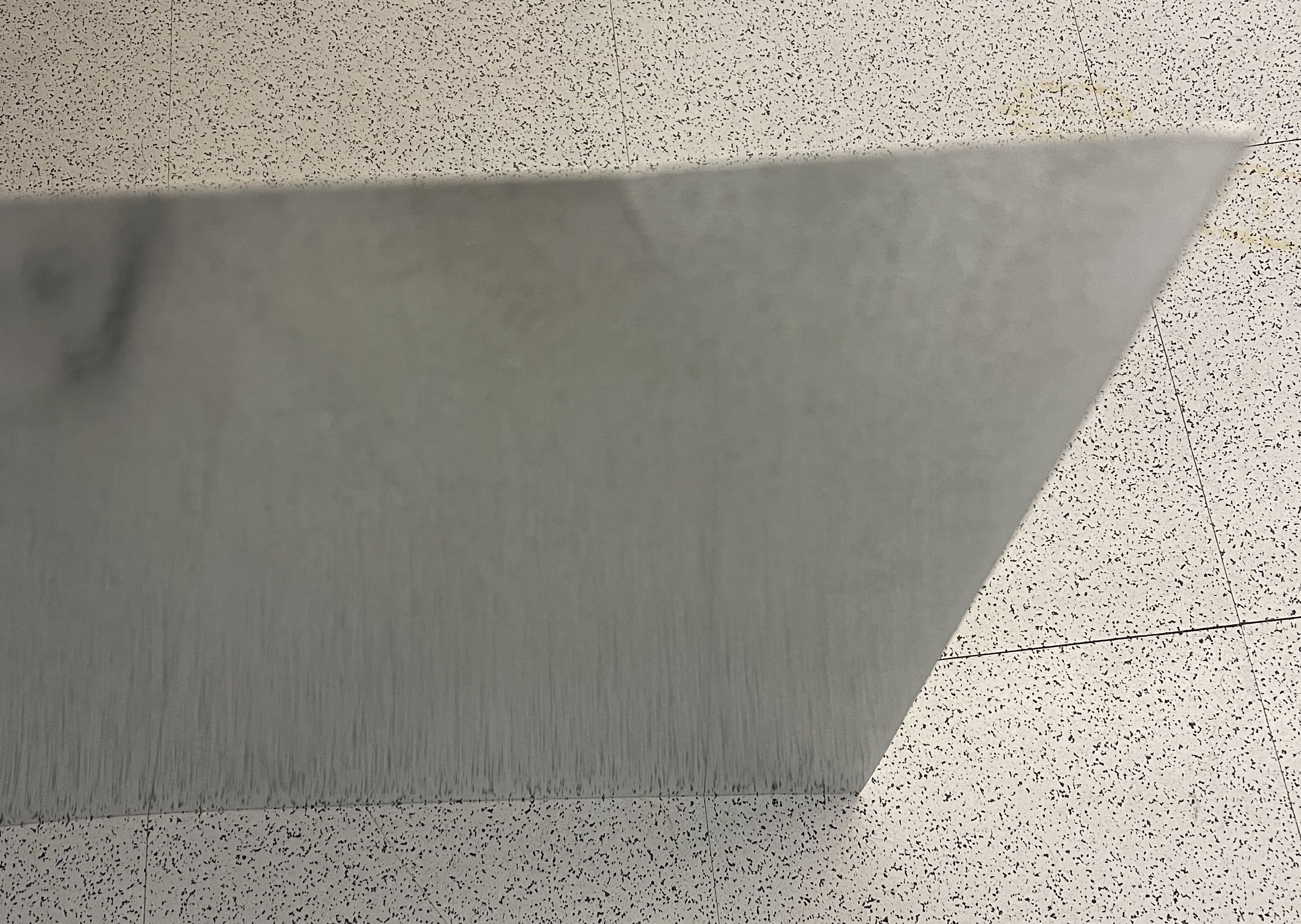}}\quad
   \subfloat[Transparent reflector. ]{\includegraphics[width=.23\textwidth,height=.18\textwidth]{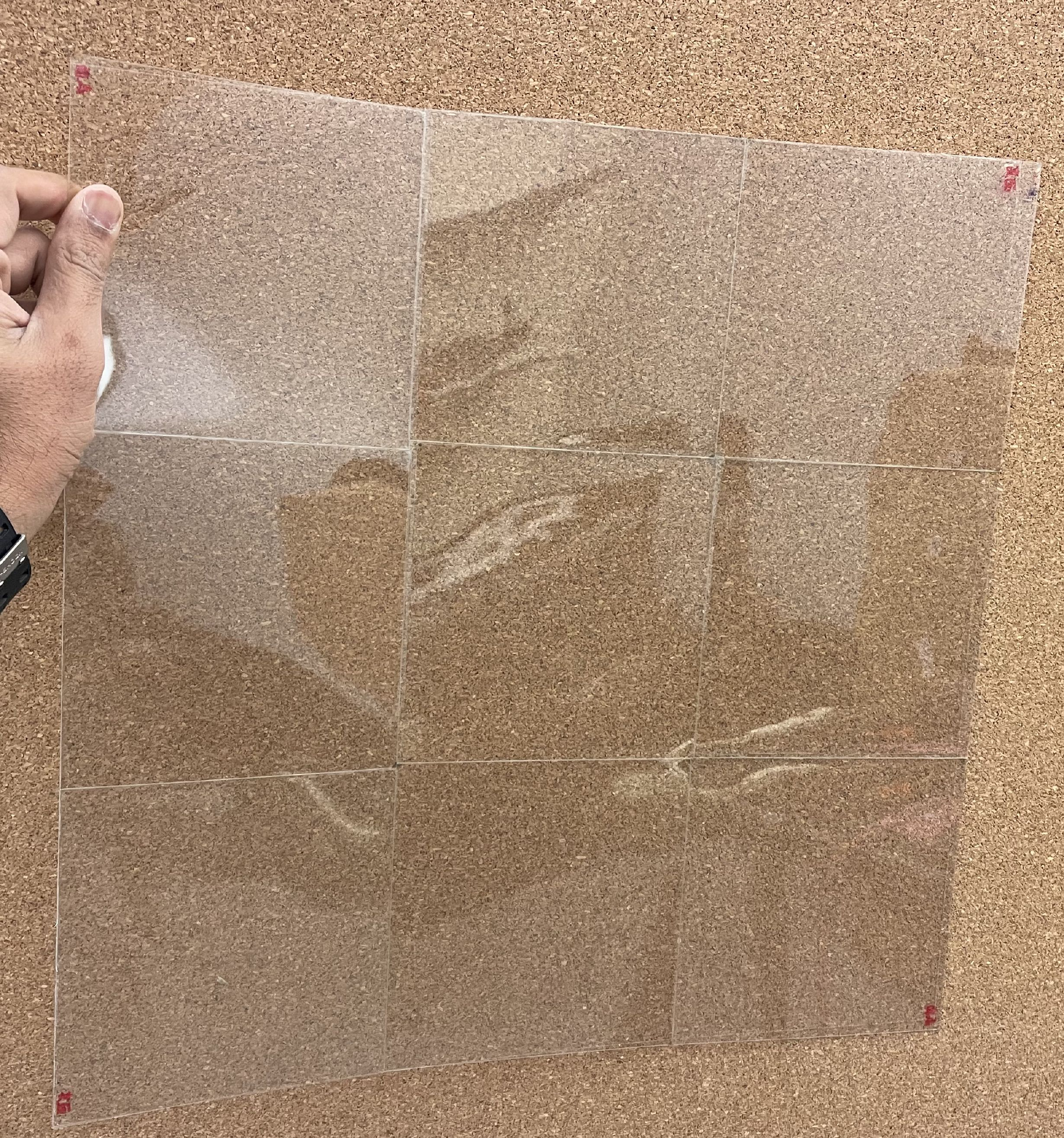}}
   \caption{Common indoor materials and transparent reflector.}
   \label{fig:Materials}
\end{figure}

In this work, along with characterizing the transparent reflector radio properties, we also compare it against other commonly used indoor materials such as ceiling tile, clear glass, drywall, plywood, and metal {(stainless steel)}, which are shown in Fig.~\ref{fig:Materials}. Each of these materials has varying dimensions quantified in Table~\ref{Table_Sim}. In particular, these thicknesses play a pivotal role in penetration loss characterization. A normalized penetration loss is also presented in Section~\ref{Sec:Results_PL} which quantifies the attenuation per unit thickness. Further, the reflection characteristics of the transparent reflector will only be compared against the metal, as it is an ideal reflector; other materials are ignored due to space constraints and we will include in our future work. In this context, we also like to point the reader to our prior work~\cite{kairui} where we have characterized penetration loss alone (and not reflection) for the above-discussed materials except the metal and transparent reflector; a key contribution of this~work.

\begin{table}[b!]
\centering
\renewcommand{\arraystretch}{1.2}
\begin{tabular}{P{2.5cm}|P{2.25cm}|P{2.25cm}}
\hline
Material& Dimension (in cm) & Thickness (in mm)  \\
\hline 
Ceiling Tile &121.92 x 60.9& 14.34 \\ \hline
Clear Glass &76.2  × 91.4  & 2.23 \\ \hline
Drywall &121.92  x 243.8 & 16.30\\ \hline
Plywood &121.92 x 57.91 & 13.20 \\ \hline
Metal &60.96  x 91.44  & 0.50\\ \hline
Transparent reflector &41.91 x 41.91 & 0.42 \\ \hline
\end{tabular}
\caption{Materials used in the measurements.}\label{Table_Sim}
\end{table}

\begin{figure*}[t!]
    \centering
    {\subfloat[ ]{\includegraphics[trim={0 {.05\textwidth} 0 0},clip,width=.33\textwidth,height=.21\textwidth]{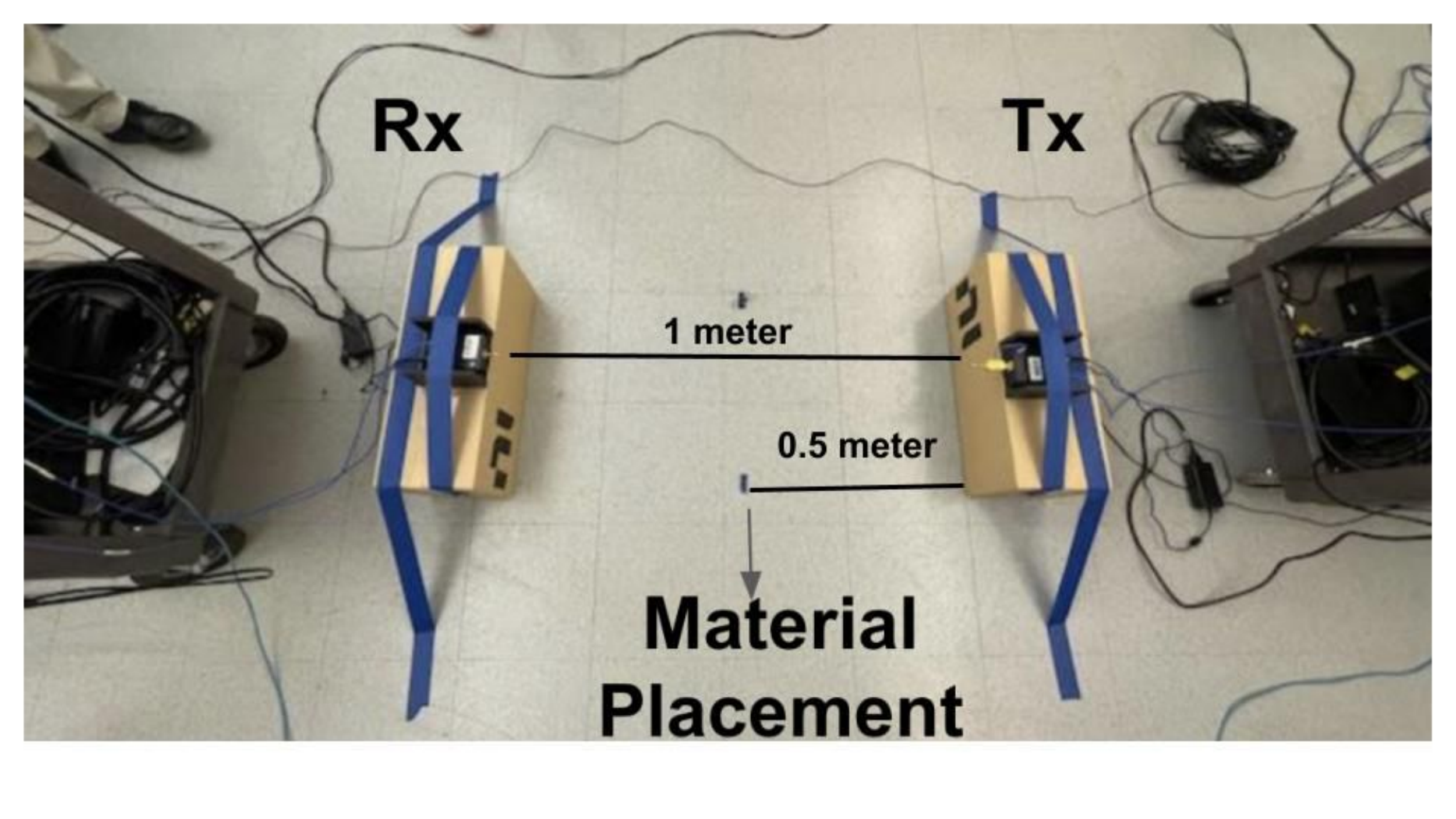}}}
    \quad
    {\subfloat[ ]{\includegraphics[width=.33\textwidth,height=.21\textwidth]{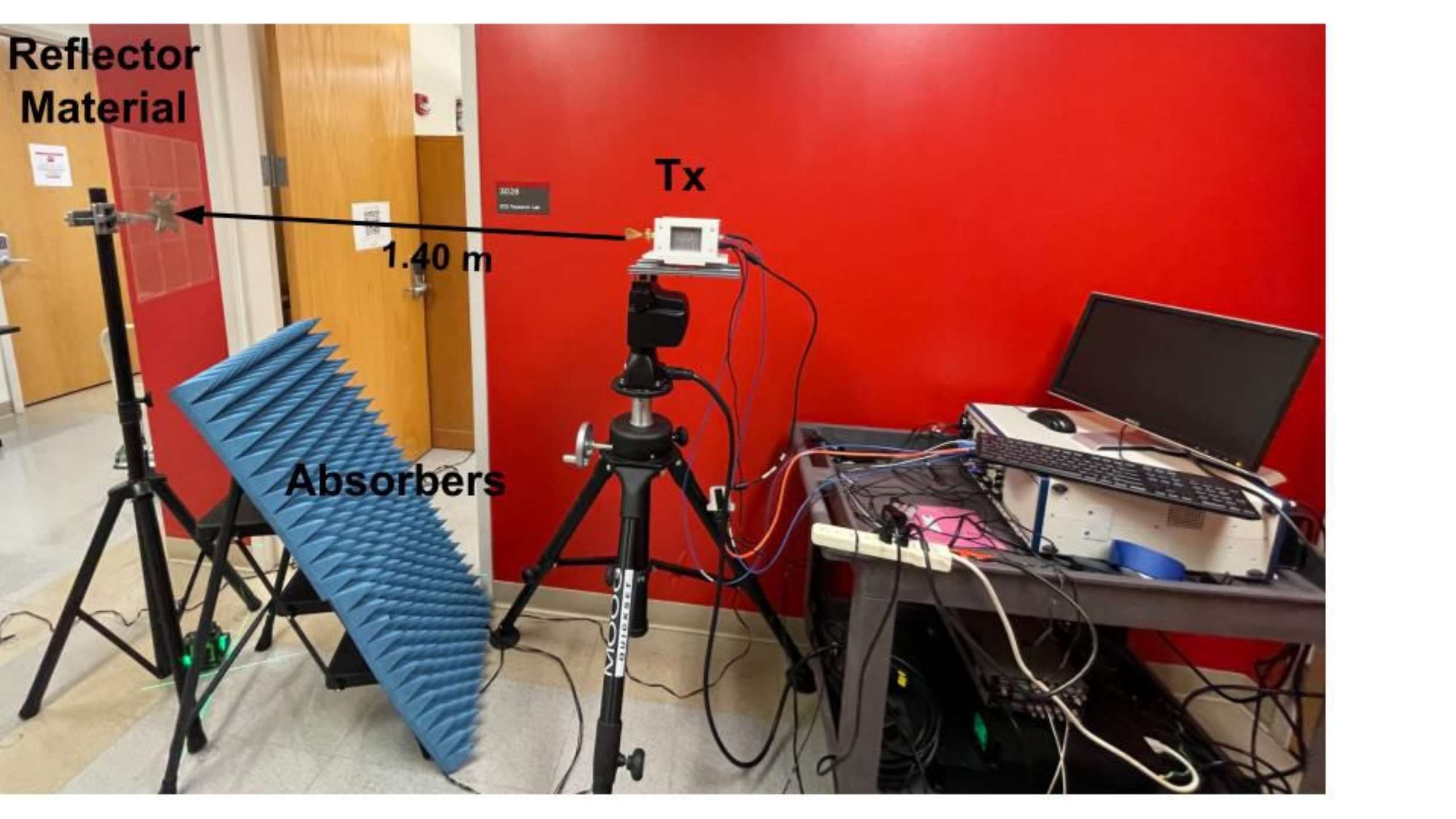}}}
    \hspace{-.2cm}
    {\subfloat[ ]{\includegraphics[trim={0 {.225\textwidth} 0 0},clip,width=.31\textwidth,height=.205\textwidth]{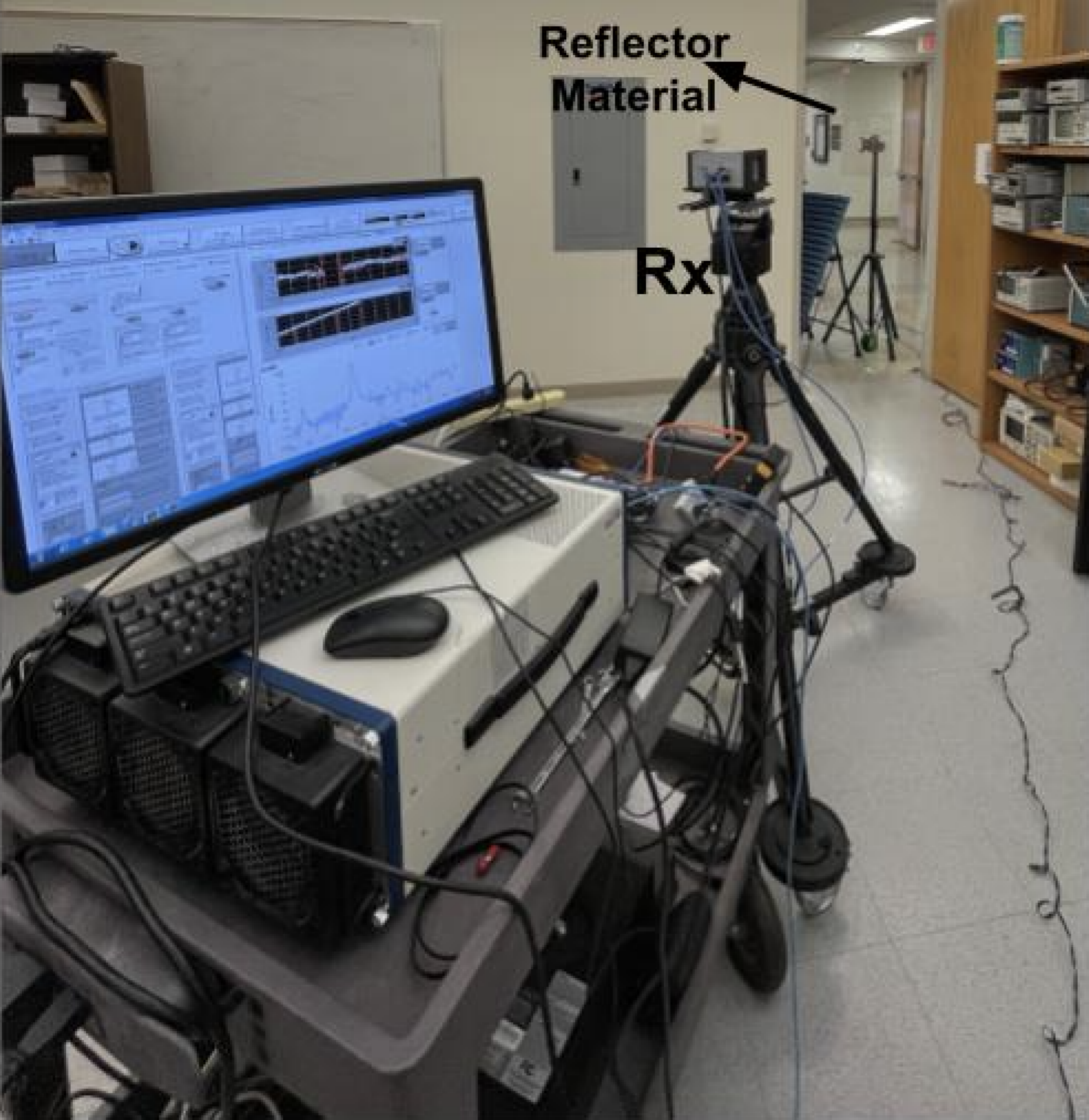}}}        \caption{{Experimental setup for characterizing (a) Penetration loss (b) Reflection: Open door measurement setup with transmitter and reflector  (c)  Reflection: Open door measurement setup with receiver.}}
        \label{fig:Exp_Setup}
        \vspace{-0.65cm}
\end{figure*}

\subsection{Channel Sounder and Equipment Details}\label{Sec:Equipment_Details}
All the measurements were conducted using a channel sounder based on NI's PXI platform~\cite{NI_Webpage} that consists of~PXIe-~1085 transmitter/receiver (Tx/Rx) chassis, Tx/Rx radio heads, and a FS725 Rubidium (Rb) clock. {All the processing is done by the PXIe-7902/8880 (FPGA/host PC inside the chassis). Further, the PXIe-3610/3630 perform the D/A and A/D conversions, whereas, the PXIe-3620 is responsible for the intermediate frequency (IF) up-conversion.}. A series of samplings and conversions are performed through the PXIe-~1085 chassis. The channel sounder transmits a 2048 bit length Zadoff-Chu sequence that is filtered by a root raised cosine (RRC) filter. The generated samples are then fed into FPGA PXIe-7902 that undergo digital to analog conversion at PXIe-3610 converter at a sampling rate of $fs$ = 3.072 GS/s. The base-band signal is then up-converted to the IF signal in the PXIe-3620 module. For 28 GHz and 39 GHz, an IF of 10.56 GHz is used. The 120 GHz and 144 GHz uses an IF of~12.6~GHz. The Tx/Rx radio heads further up-convert the IF signals to high-frequency radio signals.

At the 28 GHz setup, the Tx radio head used was NI mmRH-3642 and NI mmRH-3652 was the Rx radio head. An LO signal at 3.52 GHz is multiplied by three to generate IF signal at 10.56 GHz. Further, on the Rx end the high-frequency RF signal is down-converted to IF, by the radio-heads. The horn antennas SAR1725-34KF-E2 connected to the Tx and the Rx radio head have an antenna gain of 17 dBi Gain,~3~dB beam-width of 26 degrees, and 24 degrees in the E-plane and H-plane, respectively. The antenna system has a maximum linear dimension of 4.072 cm with a far-field of 30.97 cm.

Similarly, at 39 GHz, the Tx radio head used was NI mmRH-3643 and NI mmRH-3653 was the Rx radio head. IF signal generation is similar to our 28 GHz setup. The horn antennas SAR2013-222F-E2 connected to the Tx and the Rx radio head have an antenna gain of 20 dBi Gain, 3dB beam-width of 15 degrees, and 16 degrees in the E-plane and H-plane, respectively. The antenna system has a maximum linear dimension of 4.384 cm with a far-field of 50.005 cm.

At 120 GHz and 144 GHz, the Tx radio head used was VDI WR6.5CCU and VDI WR6.5CCD was the Rx radio head. In this case the LO generated at 4.2 GHz is multiplied by 3 to generate IF signal at 12.6 GHz. Band-pass filters WR6.5BPFE116-123 and WR6.5BPFE140-148 are used for 120 GHz and 144 GHz, respectively.  The horn antennas  VDI WR-6.5 connected to the Tx and the Rx radio head has an antenna gain of 21 dBi Gain, 3 dB beam-width of 13 degrees in both E-plane and H-plane, respectively. The antenna system has a maximum linear dimension of 1.080 cm with a far-field of 11.205 cm. For more details consult~\cite{kairui}.


\section{Characterizing Penetration and Reflection}
This section describes the measurement setup used for characterizing the penetration loss of each of the materials discussed in Section~\ref{Sec:Sys_Model} and the reflection characteristics of the transparent and metal reflector. All the measurements were carried out in an indoor environment in Engineering Building~2 at NC State University, Raleigh, NC.

\subsection{Penetration Loss: Indoor Measurement}\label{SubSec:Penetration}
Fig.~\ref{fig:Exp_Setup}(a) displays the indoor measurement setup for the penetration loss experiment. The Tx and Rx radio heads were mounted on the cardboard at 10 inches from the ground and were separated by a distance of $d\approx1$ m. Both the Tx/Rx were beam aligned manually such that the radio heads center pointing to each other with a 0$^\circ$ incident angle, corresponding to the LoS scenario with no blockage/obstruction. At first, we collected the LoS measurement data (without any blockage) and identified the LoS path component (earliest amongst all the possible MPCs) and extracted the averaged-power $P_\text{Rx}^\text{LoS}$ associated with the LoS path for each measurement point, given by
\begin{equation}
P_\text{Rx}^\text{LoS} \text{[dB]} = P_\text{Tx} + G_\text{Tx} + G_\text{Rx} - PL_\text{FS},
\end{equation}
where $P_\text{Tx}, G_\text{Tx}$, and $G_\text{Rx}$ are the transmit power, the gain at the Tx, and the gain at the Rx, respectively. The term $PL_\text{FS} = 20 \log_{10}\left(\frac{4\pi d f}{c}\right)$ denote the free space path loss, with $f$ and $c$ denoting the frequency of operation and speed of light, respectively. 

\begin{figure}[t!]
    \centering
    \includegraphics[scale=.285]{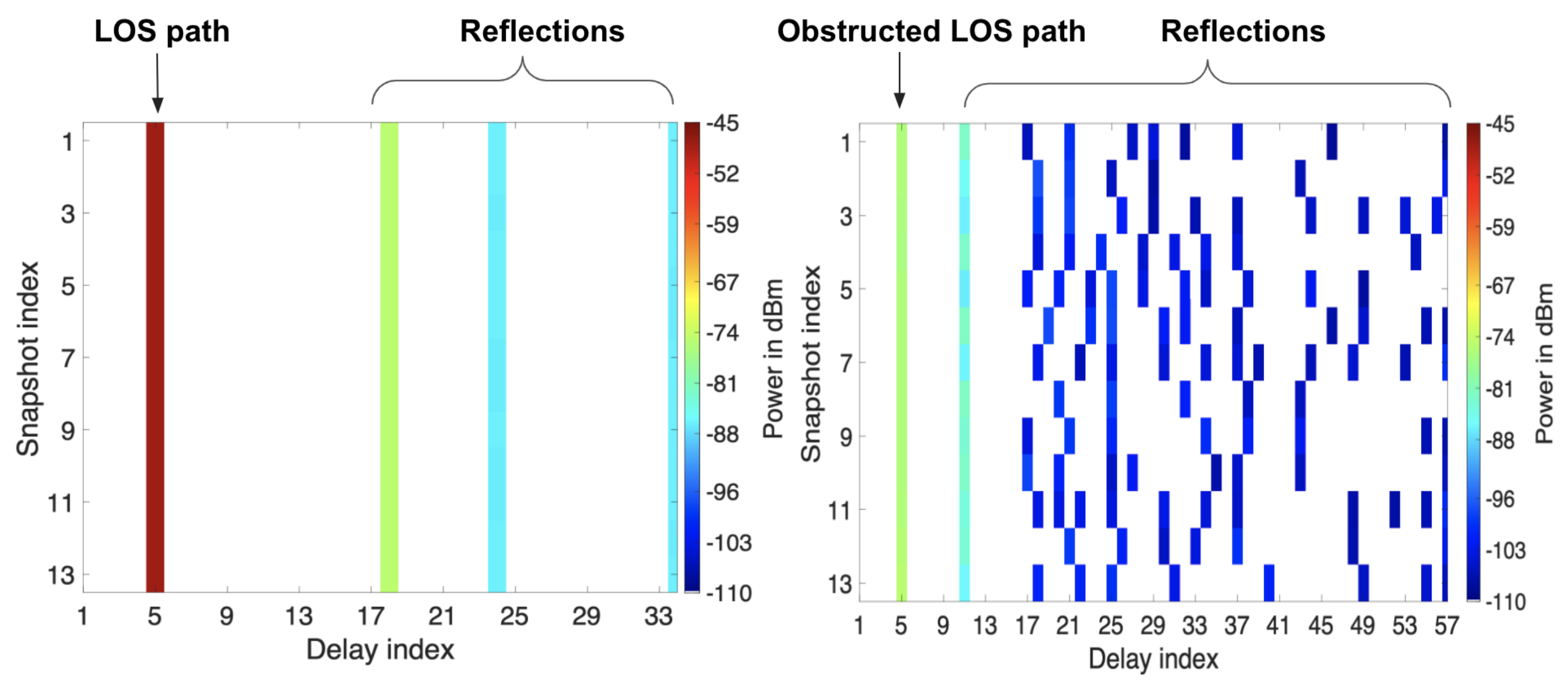}
    \caption{$L_{Pen.}$ calculation at 39 GHz: (left) LOS and (right) transparent reflector measurement. {Each delay index corresponds to a delay of 0.65 ns.}}
    \label{fig:PL_Example_39}
\end{figure}

The penetration loss (signal attenuation) through a material depends mainly on the material characteristics, frequency of operation, and the incident angle~\cite{jun2020penetration}. We characterize the penetration loss of different materials at frequencies discussed in Section~\ref{Sec:Common_Materials}. To do so, we place the material of interest midway (at 0.5 m) between the Tx and Rx at an incident angle of 0$^\circ$, as shown in Fig.~\ref{fig:Exp_Setup}(a). We leave the experiments for different incident angles as future work. {It is also worth mentioning that another interesting future work direction is the optimal deployment of transparent reflectors to enhance coverage~\cite{anjinappa2020base}.} Similar to the LoS measurement data, we collected measurement points with obstruction via the material and measured the power $P_\text{Rx}^\text{OLoS}$ associated with the obstructed LoS, given by
\begin{equation}
P_\text{Rx}^\text{OLoS} \text{[dB]} = P_\text{Tx} + G_\text{Tx} + G_\text{Rx} - PL_\text{FS} - L_\text{Pen.}.
\end{equation}
The term $L_\text{Pen.}$ denotes the penetration loss induced by the material signifying the attenuation introduced due to obstruction. Note that the signal attenuation through a material depends on the thickness of the material; see Section~\ref{Sec:Results_PL} for normalized attenuation results. Finally, the term $L_\text{Pen.}$ can be estimated by subtracting the averaged obstructed LoS power $P_\text{Rx}^\text{OLoS}$ from the averaged (unobstructed) LoS power $P_\text{Rx}^\text{LoS} $. That is, 
\begin{equation}
L_\text{Pen.} \text{[dB]} =~P_\text{Rx}^\text{LoS} - P_\text{Rx}^\text{OLoS}    . 
\end{equation}
Examples of the above phenomenon appear in Fig.~\ref{fig:PL_Example_39}. The left and right figures show the measurements obtained for the LoS and the transparent material at 39 GHz, respectively. As demonstrated, we only consider the LoS path and obstructed LoS that arrive at the identical time delay for calculating $L_\text{Pen.}$, ignoring the reflections (ground and other reflections from the background and the material itself). We also emphasize that there was no beam spillage problem as the size of the materials was larger than the beam area~\cite{ntontin2020reconfigurable}.

\subsection{Reflection: Open Door Measurement}\label{Sec:Open_Door_Exp}
In this sub-section, we describe the experimental setup for characterizing the reflection property of the material. The measurement setup is a practical communication scenario (open-door indoor environment), where the Tx is outside the room, and the Rx is inside the room, with the LoS path obstructed by the wall. To extend the coverage, we deploy passive reflectors to provide the NLoS link between the Tx/Rx, as illustrated in Fig.~\ref{fig:Exp_Setup}(b). For the current experiment, we compare the transparent reflector against the metal reflector alone (as metal is the perfect conductor). We emphasize that, unlike the previous experiment with small distances, beam spillage would be an issue here due to larger distances between the Tx/reflector/Rx, respectively; consult~\cite{ntontin2020reconfigurable} for details. Thus, reflectors of different sizes would have different beam spillage ratios. For a fair comparison, the metal reflector size was trimmed down to a similar size as the transparent reflector (thickness is the same as in Table~\ref{Table_Sim}). 
Even though the beam spillage phenomenon exists, it would be the same for both cases. Further, the reflectors were mounted on a tripod such that it forms 45$^\circ$ incident and reflected angle between the Tx/Rx, respectively. The distance between Tx/reflector center and reflector center/Rx was 1.40 m and 3.43 m, respectively. Further, the {broadband pyramidal absorbers, each of dimension $24^{'} \times 24^{'} \times 4^{'}$,} were deployed near the tripod to avoid unnecessary reflections from the tripod.

In this experiment, both the Tx and the Rx are equipped with a rotating gimbal to measure the received power from different directions. This helps in characterizing (to some extent) the diffuse scattering phenomenon that happens in radio bands~\cite{ozdogan2019intelligent}. We chose the gimbal rotation range of 90$^\circ$ to -90$^\circ$ at the Tx and 140$^\circ$ to -140$^\circ$ at Rx, respectively. A 5$^\circ$ resolution was maintained at both Tx/Rx gimbal, this separation assured a trade-off between the complete coverage and experiment run-time. We take the same approach as in the previous section to characterize the reflectivity of metal and transparent reflectors. Unlike the simple analytical characterization of the penetration loss, the reflection power depends on multiple factors such as the incident/departure angles formed between the reflector and Tx/Rx, the orientation of the reflector, the beam area formed on the reflector, and the area of the reflector~\cite{ntontin2020reconfigurable,anjinappa2020base}. Thus, we avoid the theoretical framework in this study and calculate the total received power. This simplifies the way to characterize the material reflectivity, as presented in Section~\ref{Sec:Results_Reflection}. 

\begin{figure}[b!]
\vspace{-.25cm}
    \centering
    {\subfloat[Absolute penetration loss.]{\includegraphics[width=.54\textwidth,height=.3\textwidth]{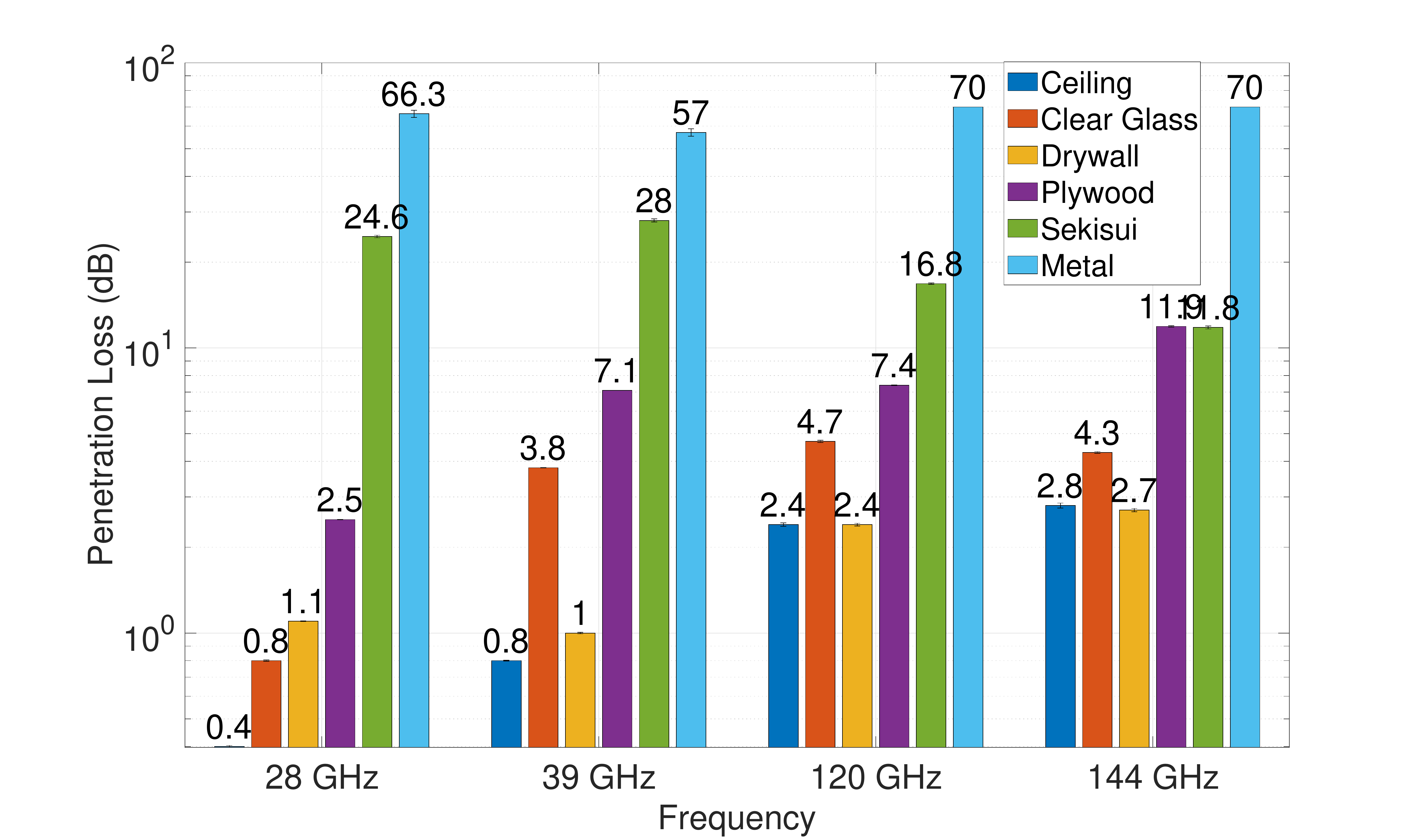}}}
    \\
    {\subfloat[Normalized penetration loss (attenuation) per centimeter.]{\includegraphics[width=.54\textwidth,height=.3\textwidth]{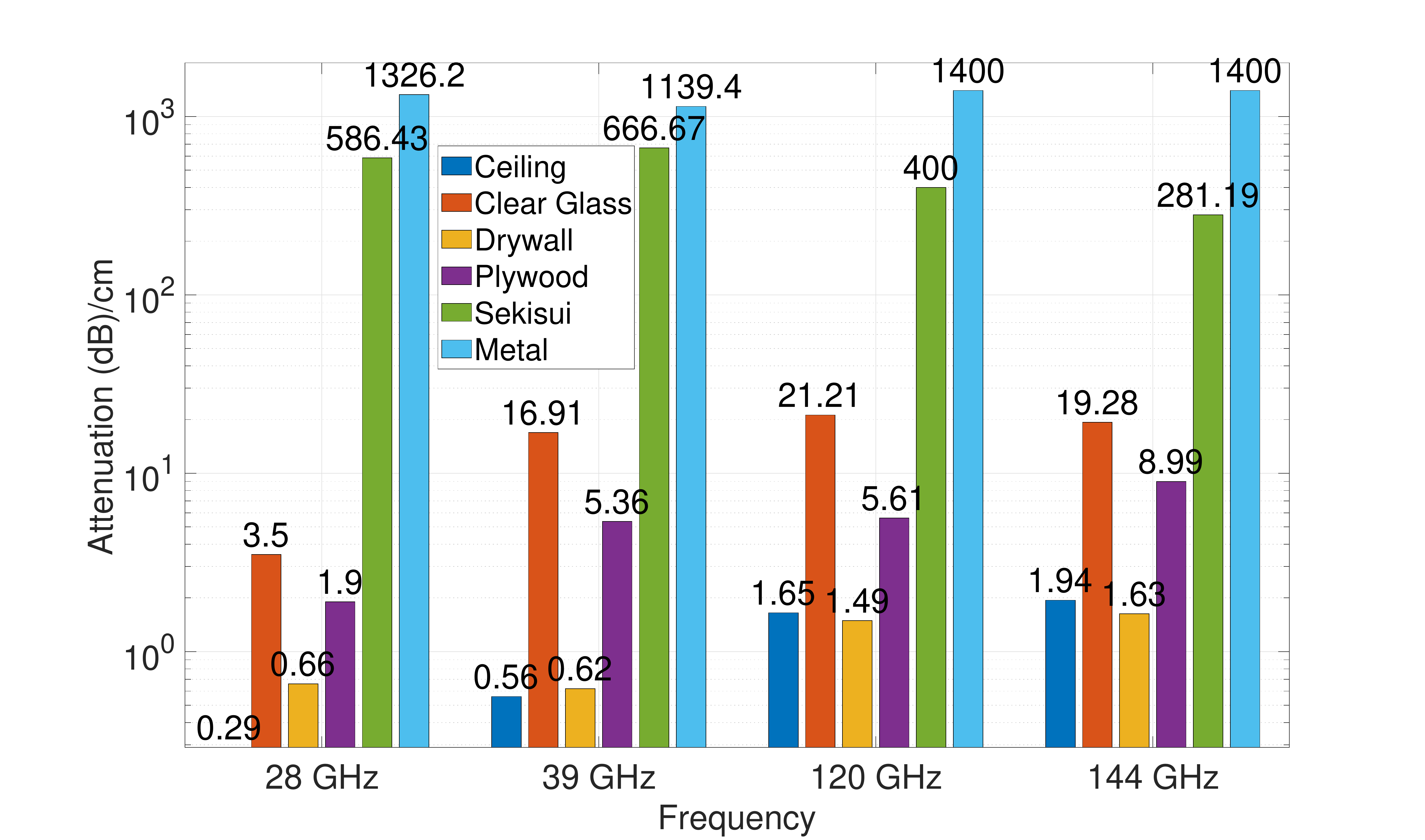}}}
        \caption{Penetration loss results at mmWave and sub-THz frequencies.}
     \label{fig:Penetration_Loss}
\end{figure}

\begin{figure*}[b!]
\vspace{-.4cm}
    \centering
     \subfloat[No reflector (no tripod).]{\includegraphics[width=.2\textwidth,height=.2\textwidth]{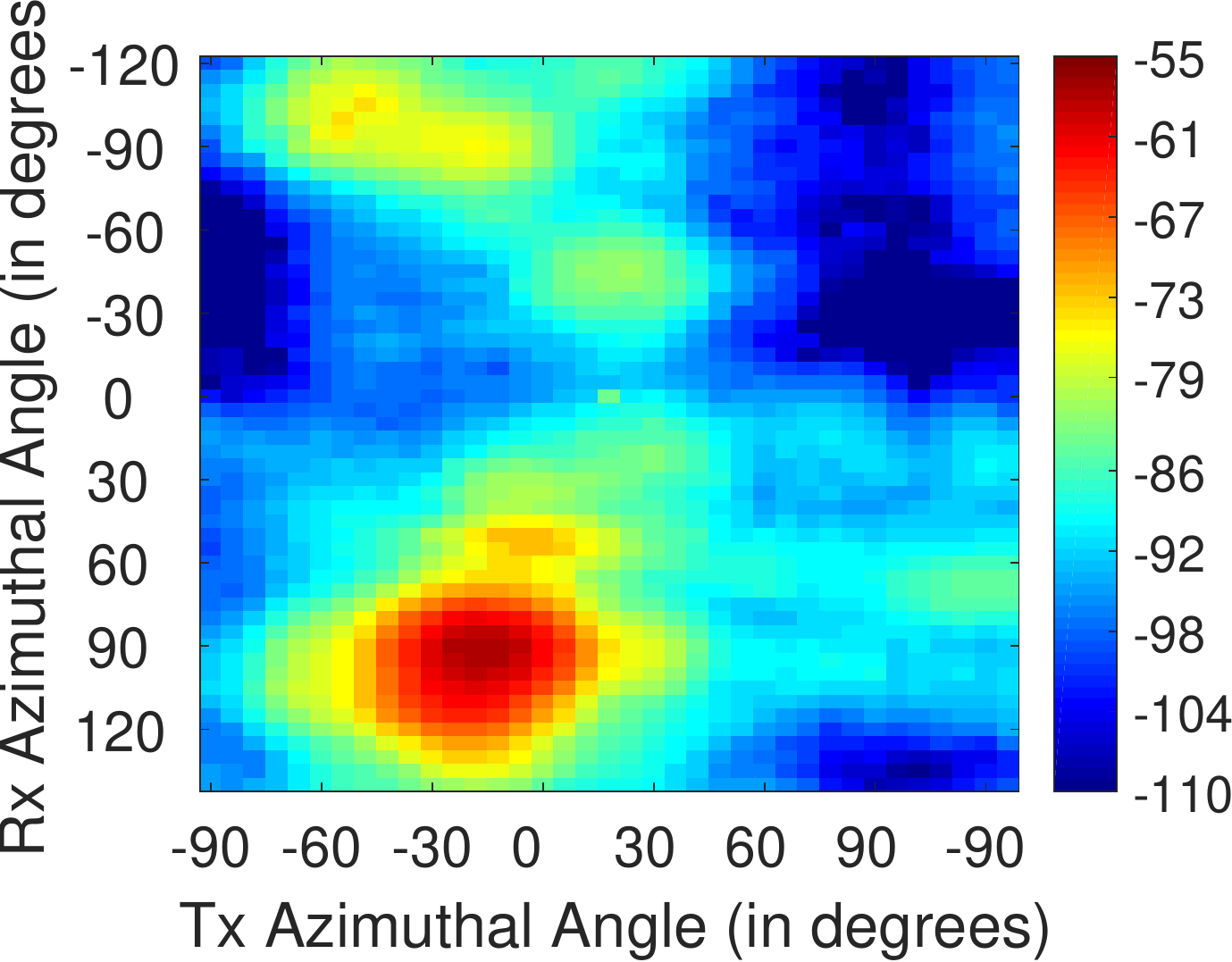}}
       \subfloat[Metal reflector.]{\includegraphics[width=.2\textwidth,height=.2\textwidth]{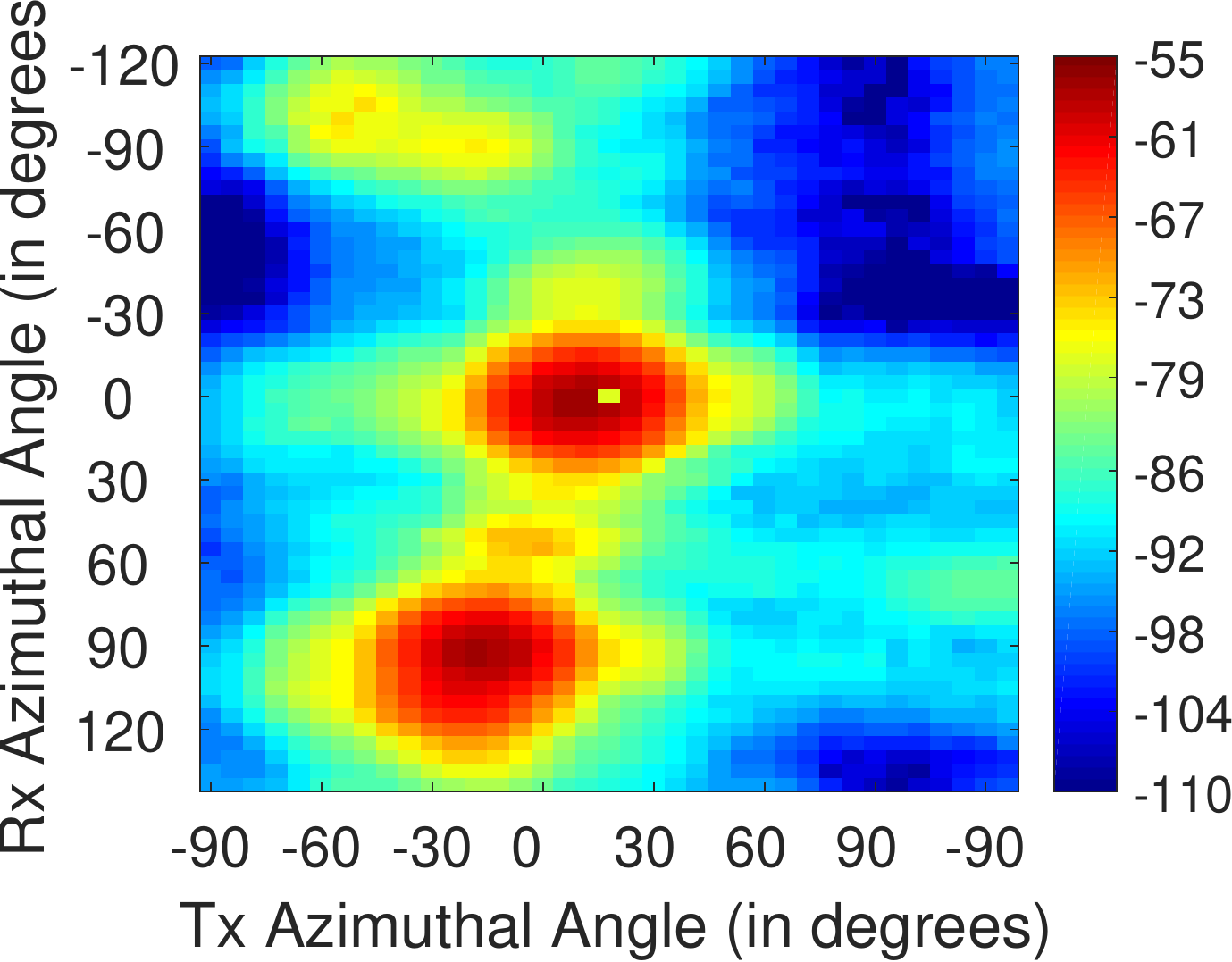}}
         \subfloat[Transparent reflector.]{\includegraphics[width=.2\textwidth,height=.2\textwidth]{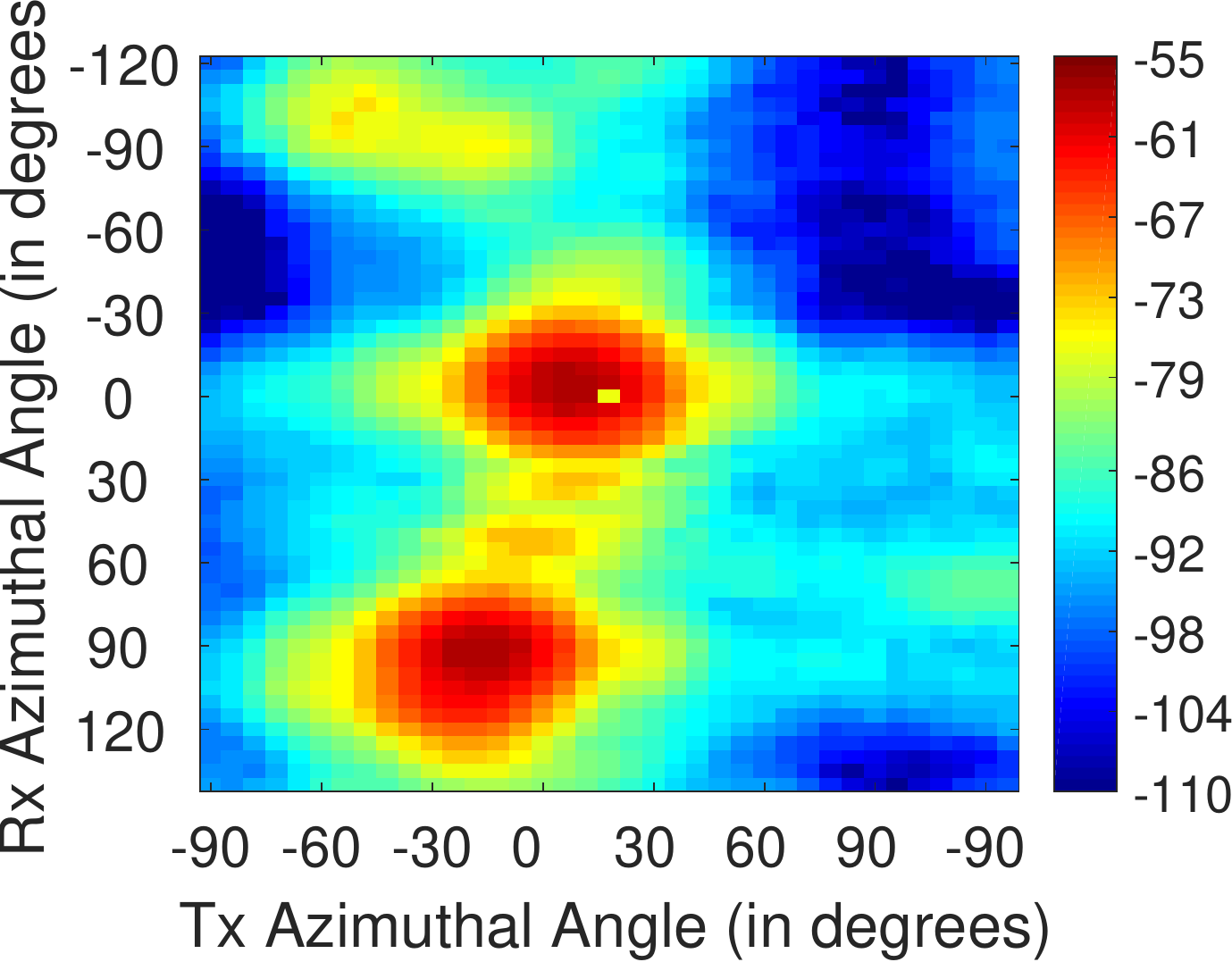}}
        \subfloat[MRC: $P_{\max}$ = 36.19 dB]{\includegraphics[width=.2\textwidth,height=.2\textwidth]{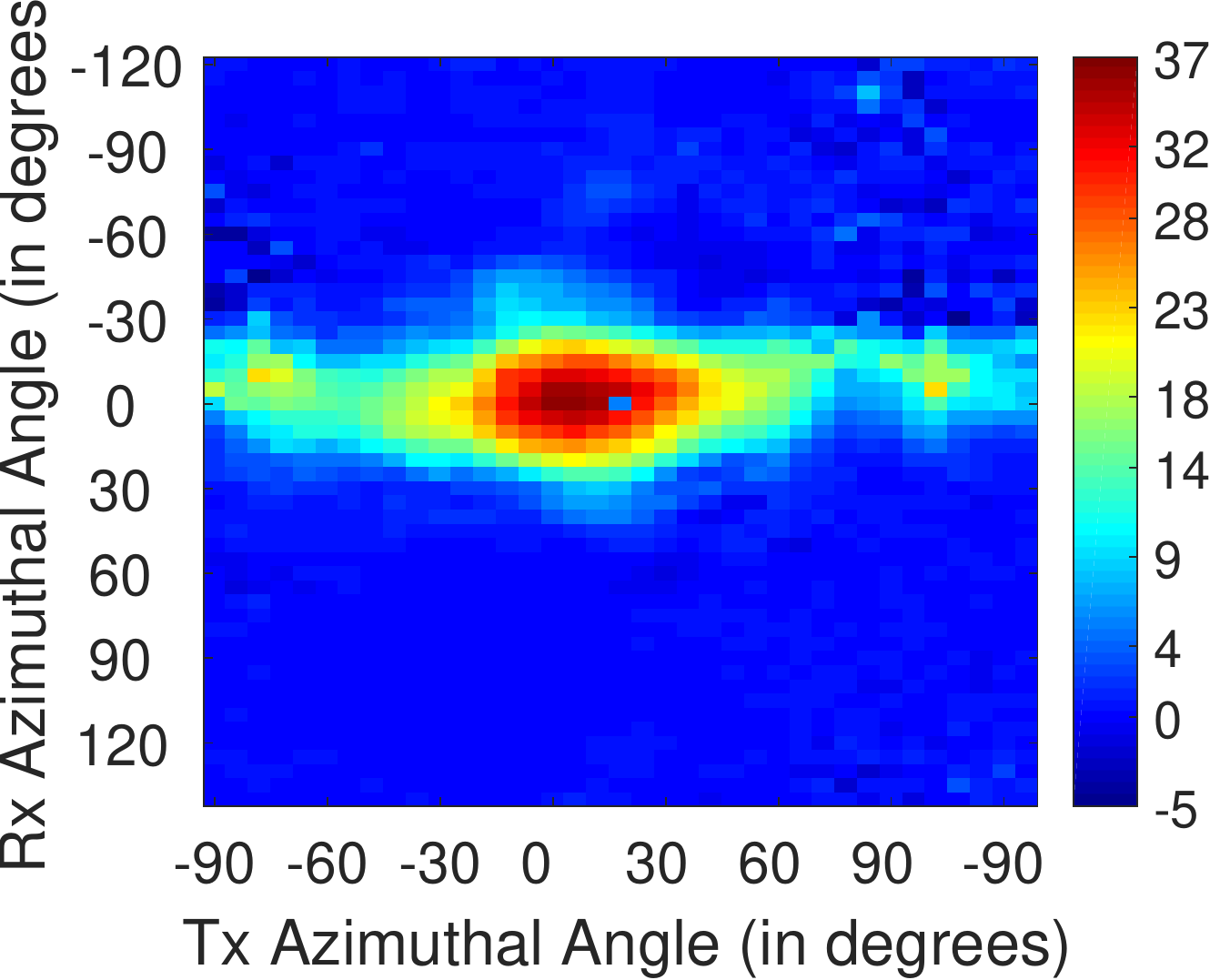}}
        \subfloat[TRC: $P_{\max}$ = 36.68 dB.]{\includegraphics[width=.2\textwidth,height=.2\textwidth]{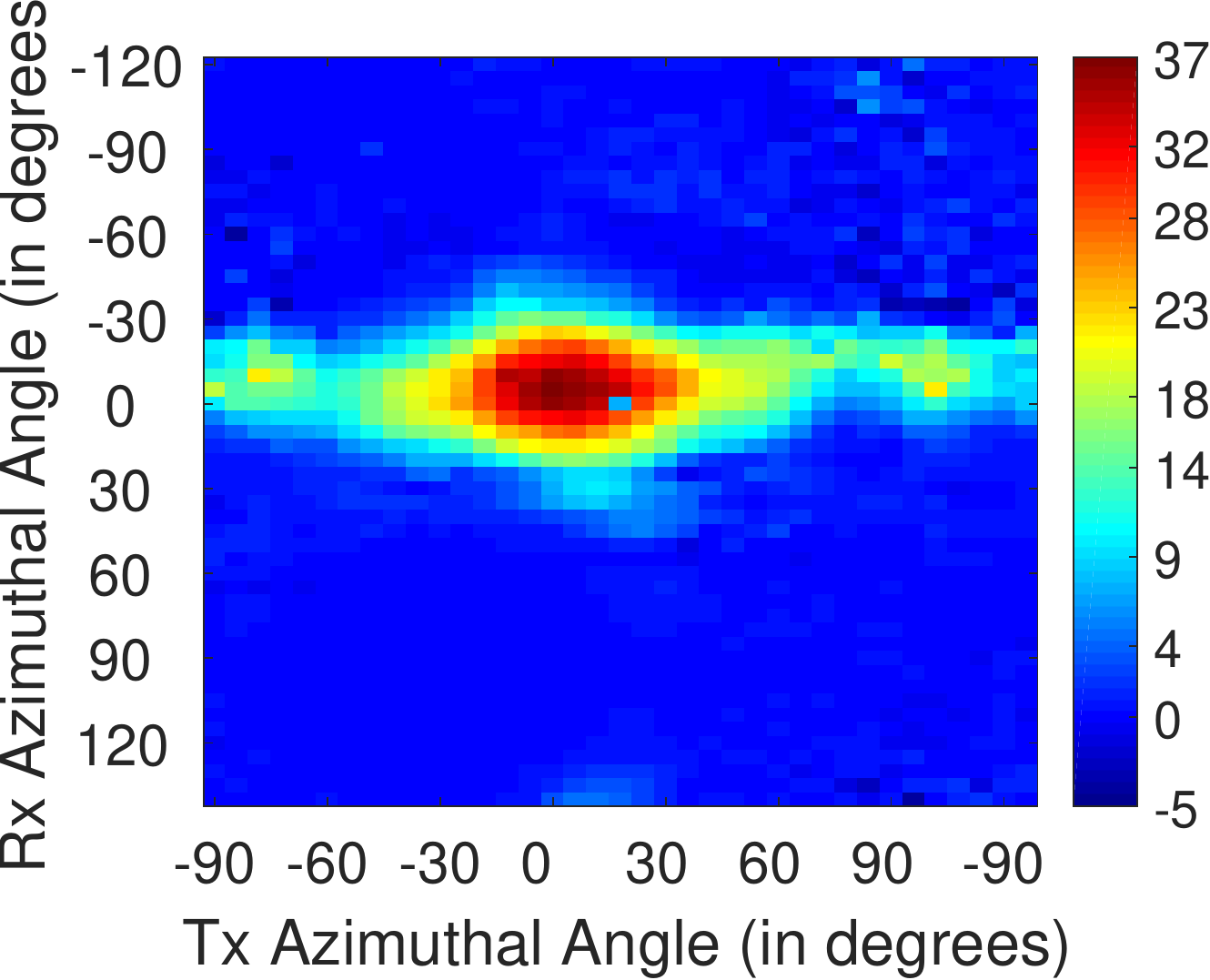}}
        \\
        \subfloat[No reflector (no tripod).]{\includegraphics[width=.2\textwidth,height=.2\textwidth]{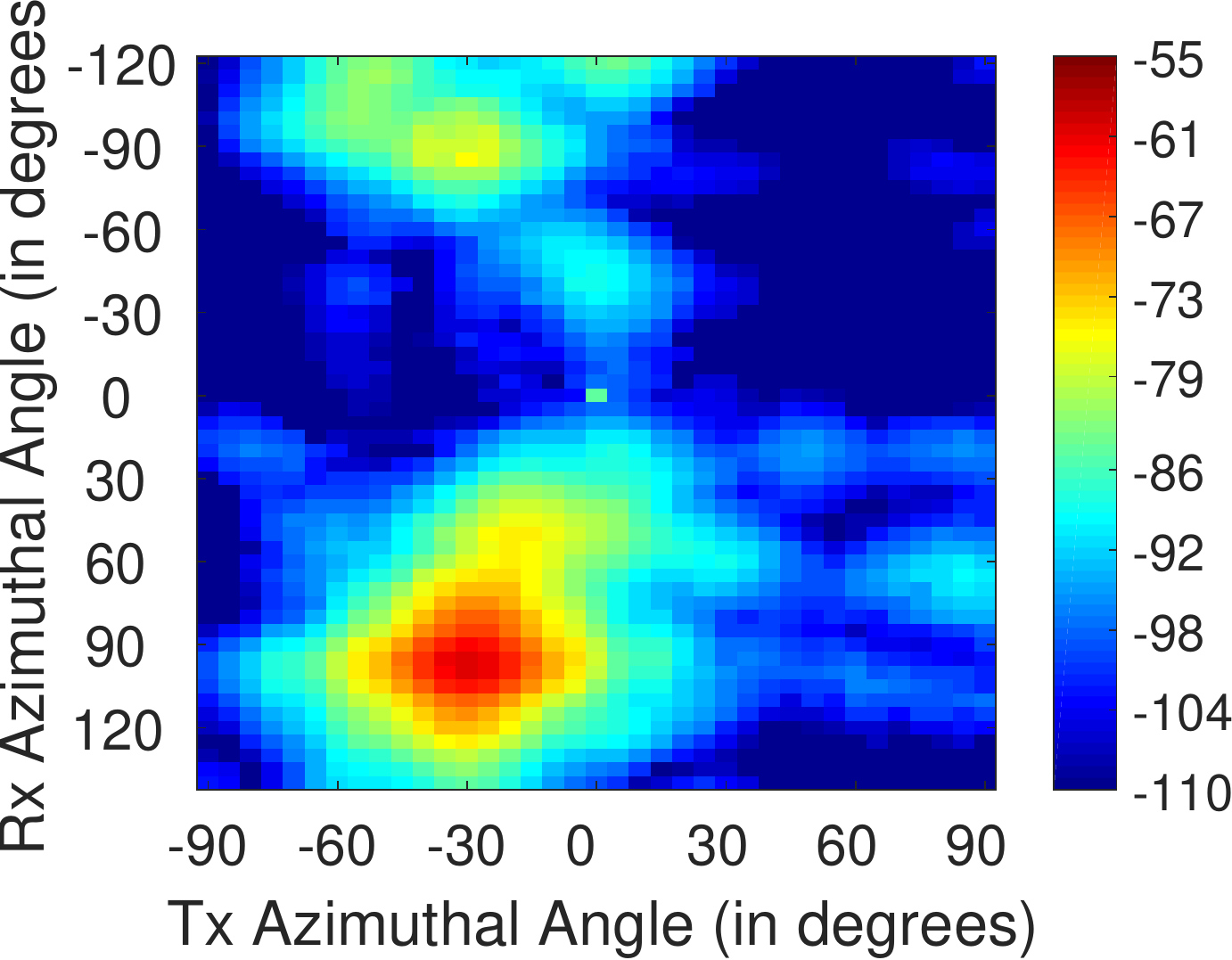}}
       \subfloat[Metal reflector.]{\includegraphics[width=.2\textwidth,height=.2\textwidth]{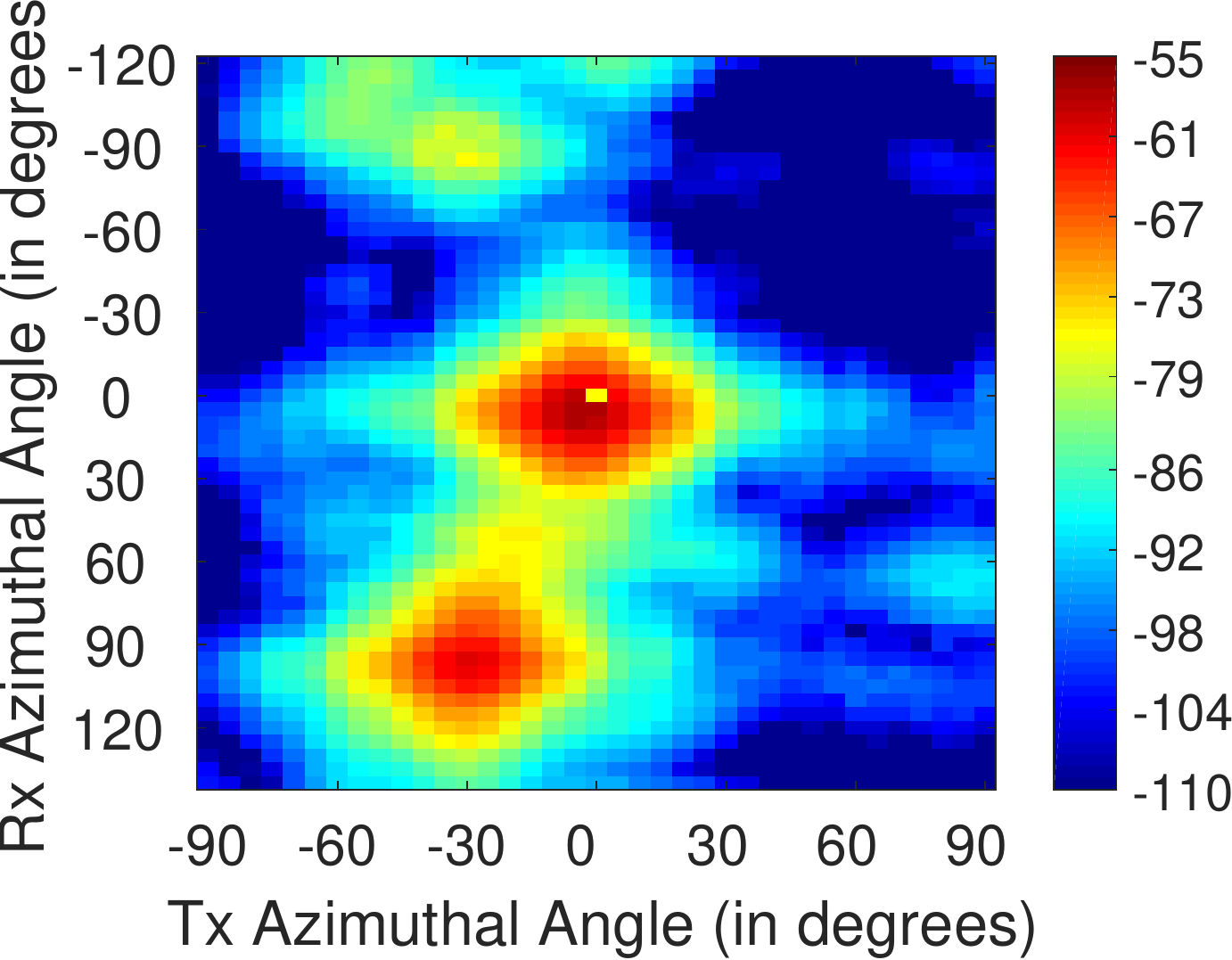}}
         \subfloat[Transparent reflector.]{\includegraphics[width=.2\textwidth,height=.2\textwidth]{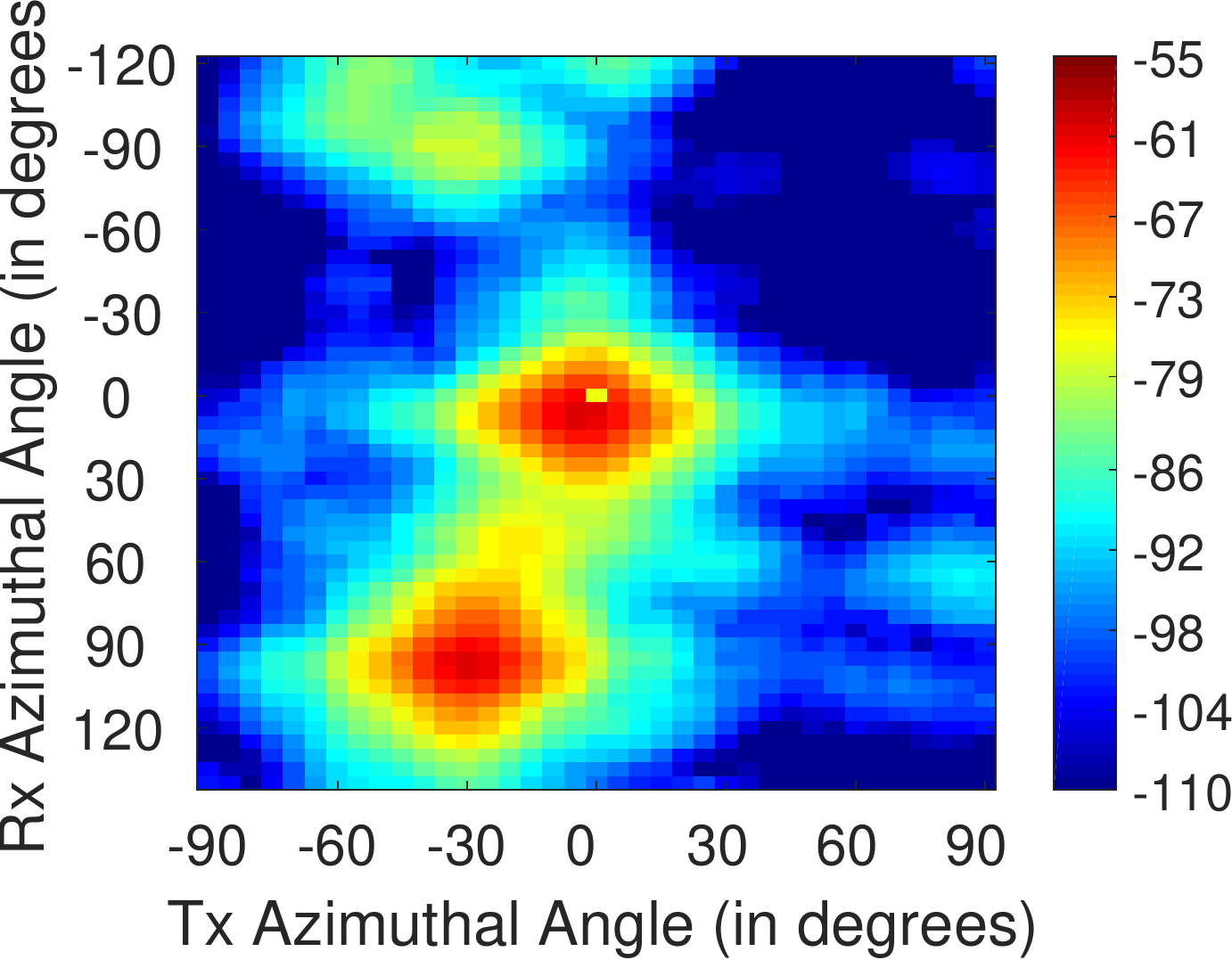}}
        \subfloat[MRC: $P_{\max}$ = 45.71 dB.]{\includegraphics[width=.2\textwidth,height=.2\textwidth]{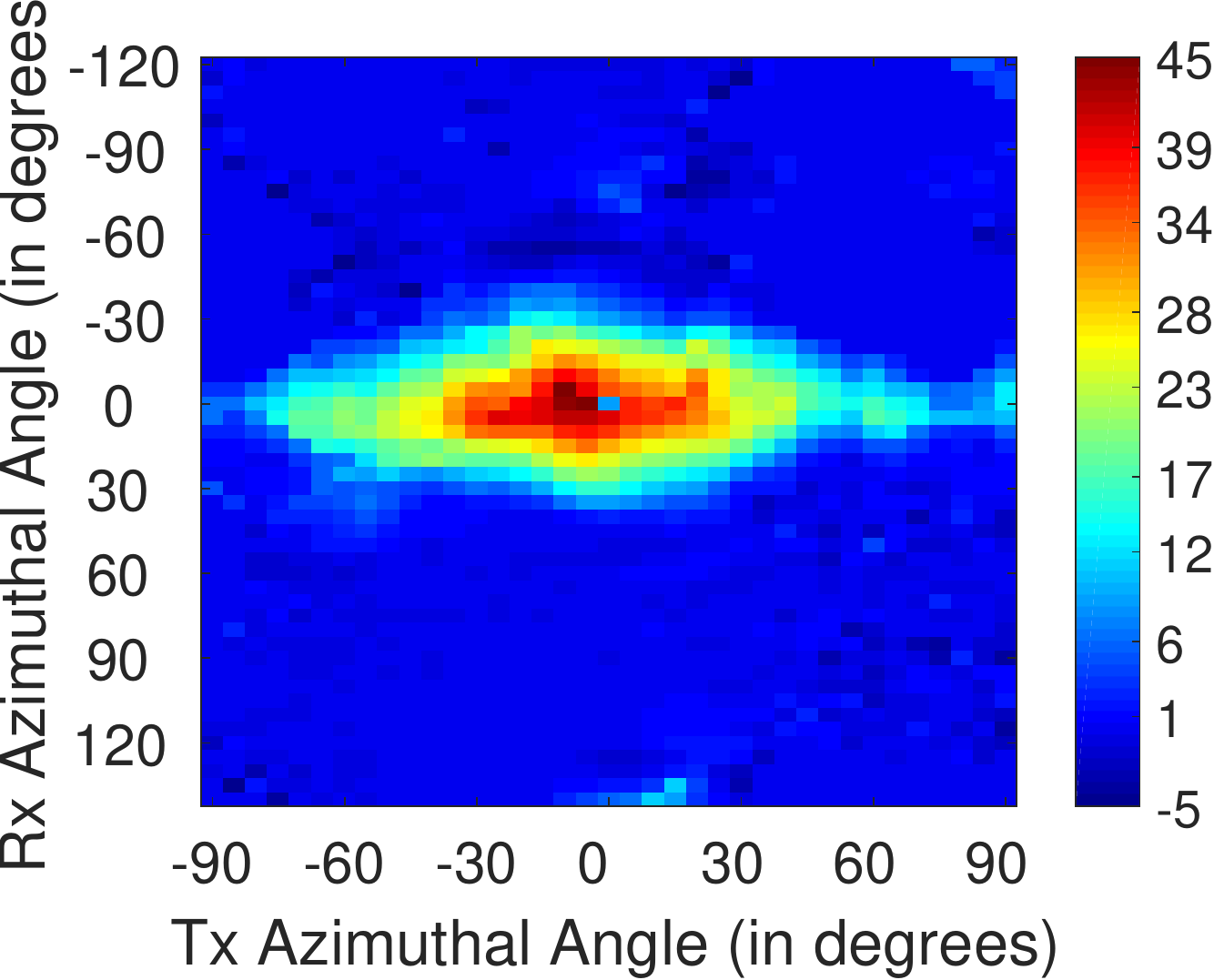}}
        \subfloat[TRC: $P_{\max}$ = 41.69 dB.]{\includegraphics[width=.2\textwidth,height=.2\textwidth]{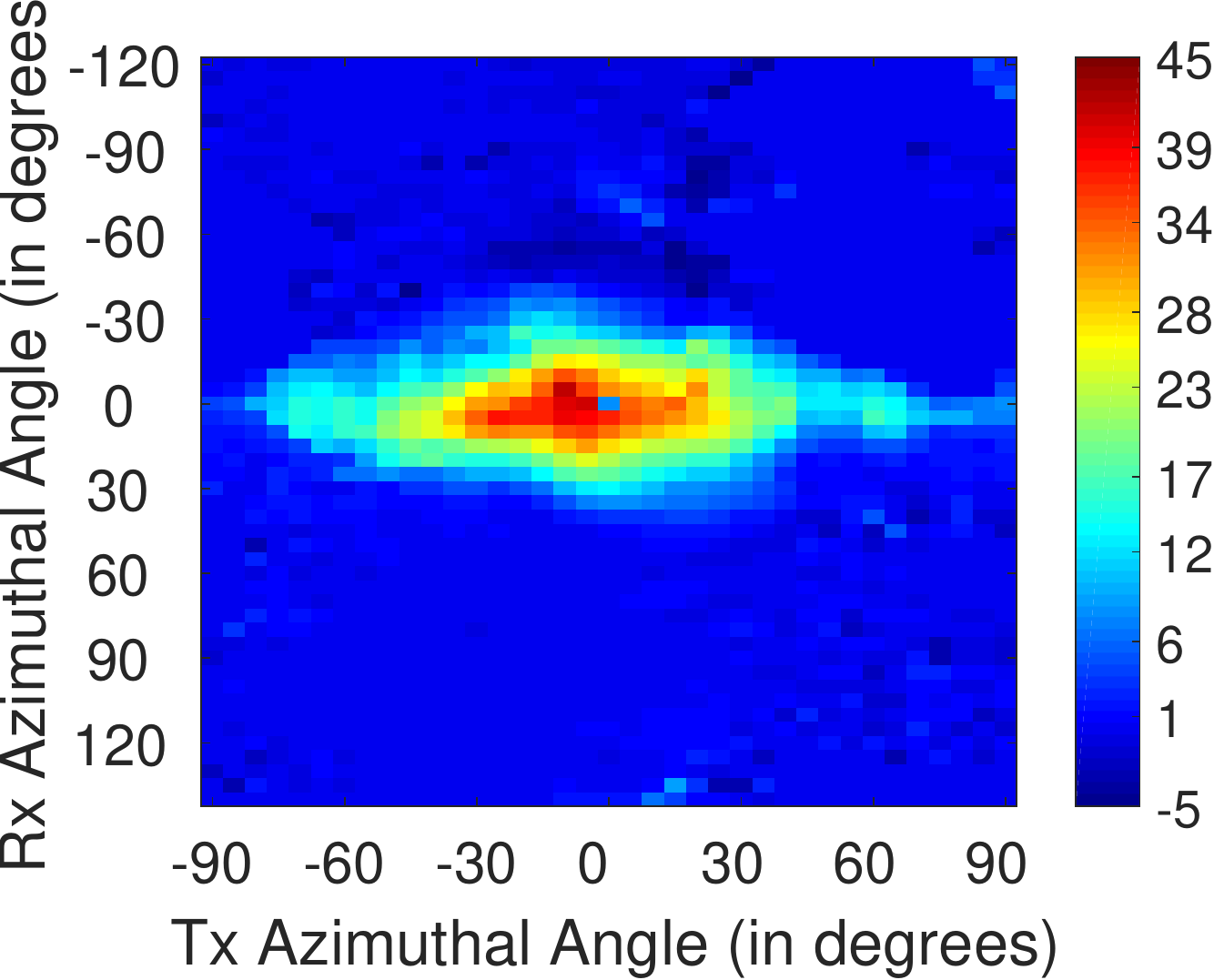}}
        \\
        \subfloat[No reflector (no tripod).]{\includegraphics[width=.2\textwidth,height=.2\textwidth]{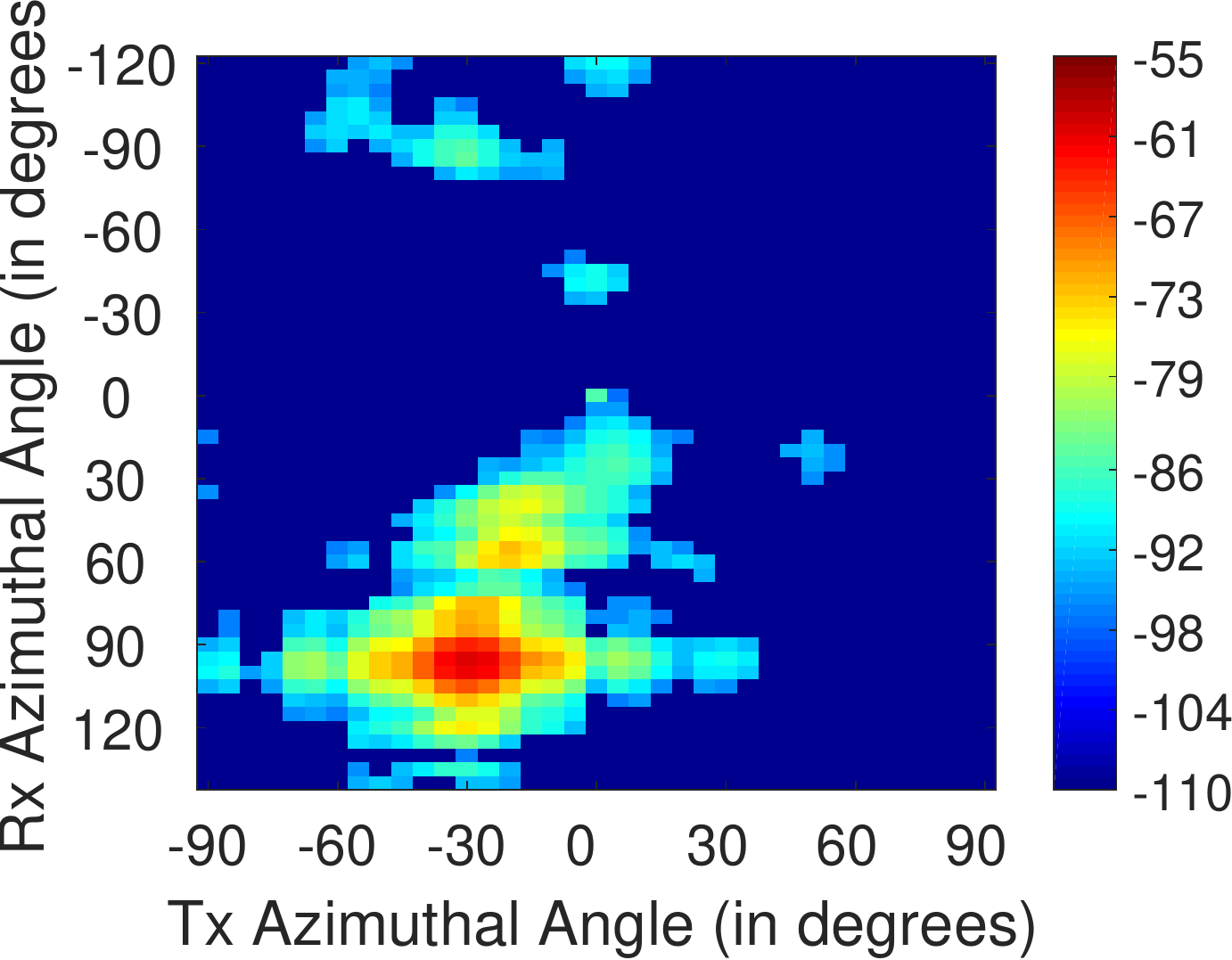}}
       \subfloat[Metal reflector.]{\includegraphics[width=.2\textwidth,height=.2\textwidth]{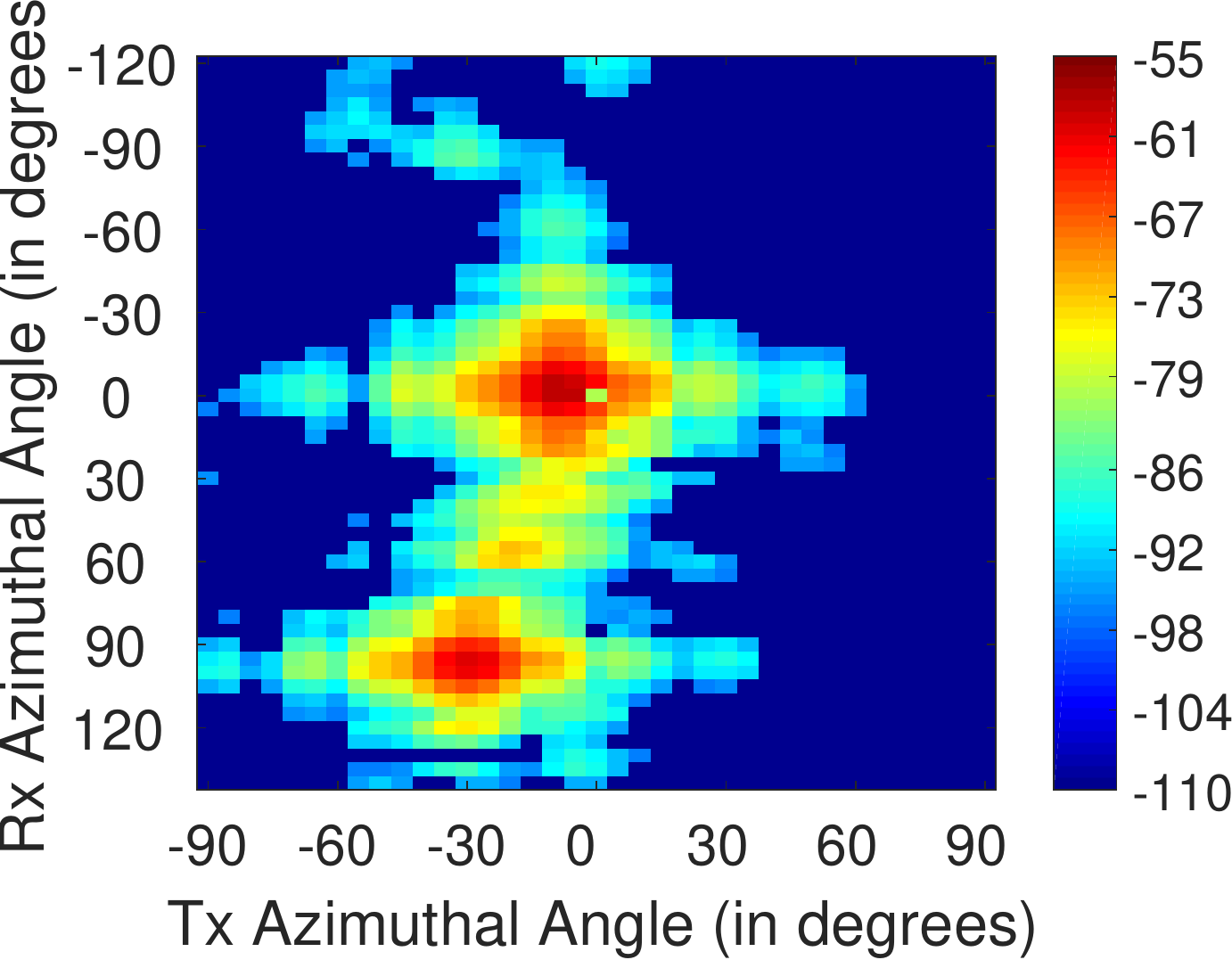}}
         \subfloat[Transparent reflector.]{\includegraphics[width=.2\textwidth,height=.2\textwidth]{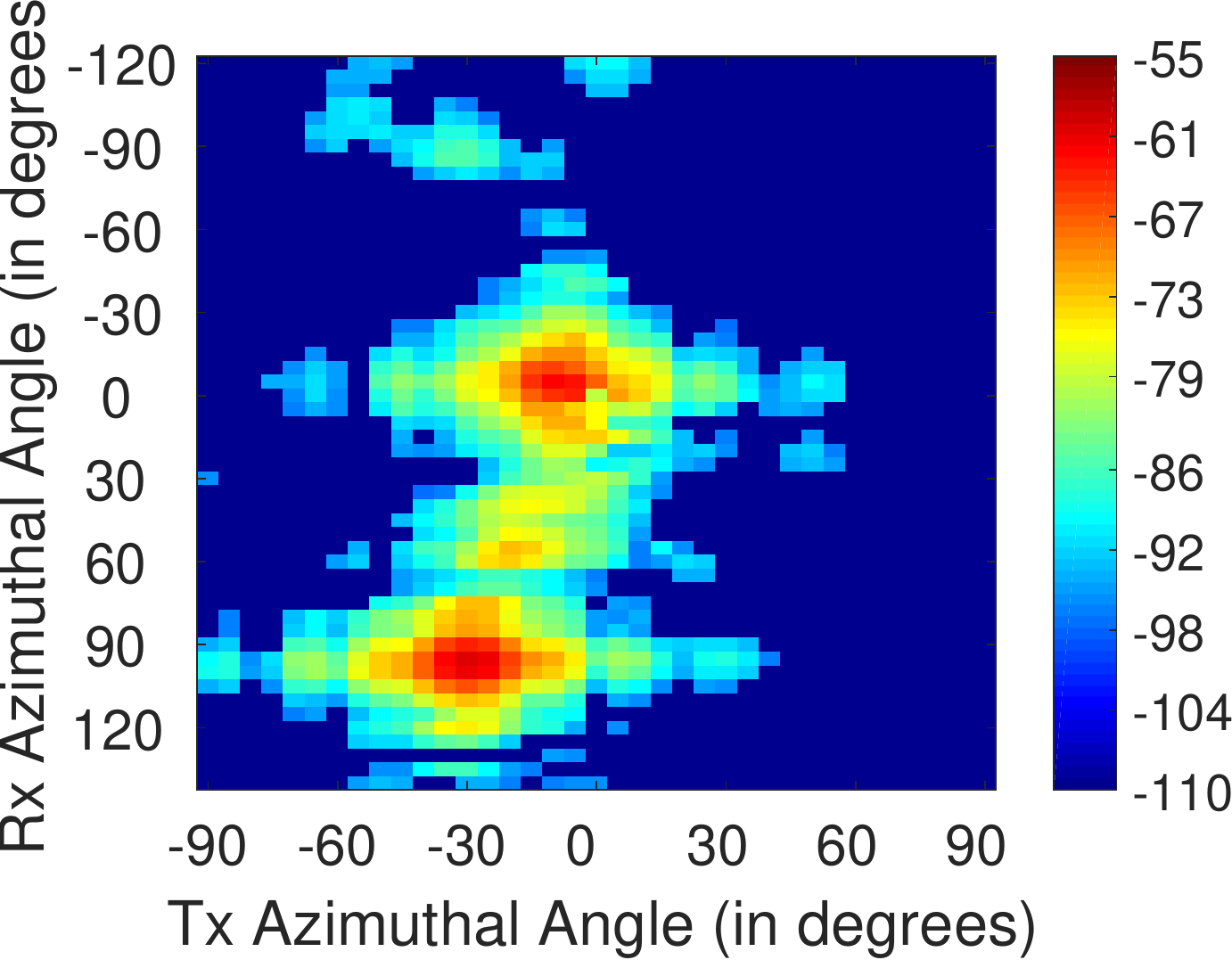}}
        \subfloat[MRC: $P_{\max}$ = 51.15 dB. ]{\includegraphics[width=.2\textwidth,height=.2\textwidth]{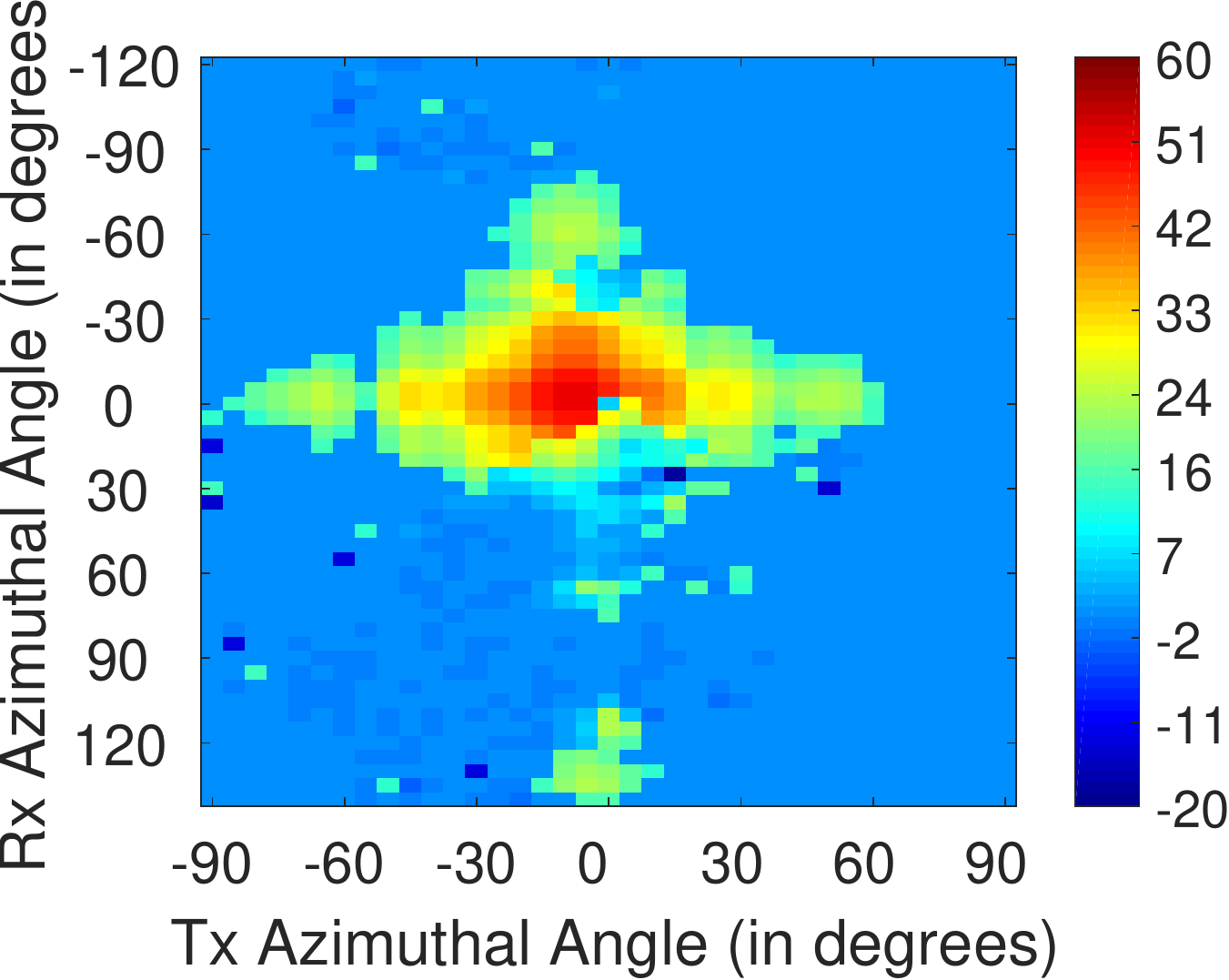}}
        \subfloat[TRC: $P_{\max}$ = 56.58 dB.]{\includegraphics[width=.2\textwidth,height=.2\textwidth]{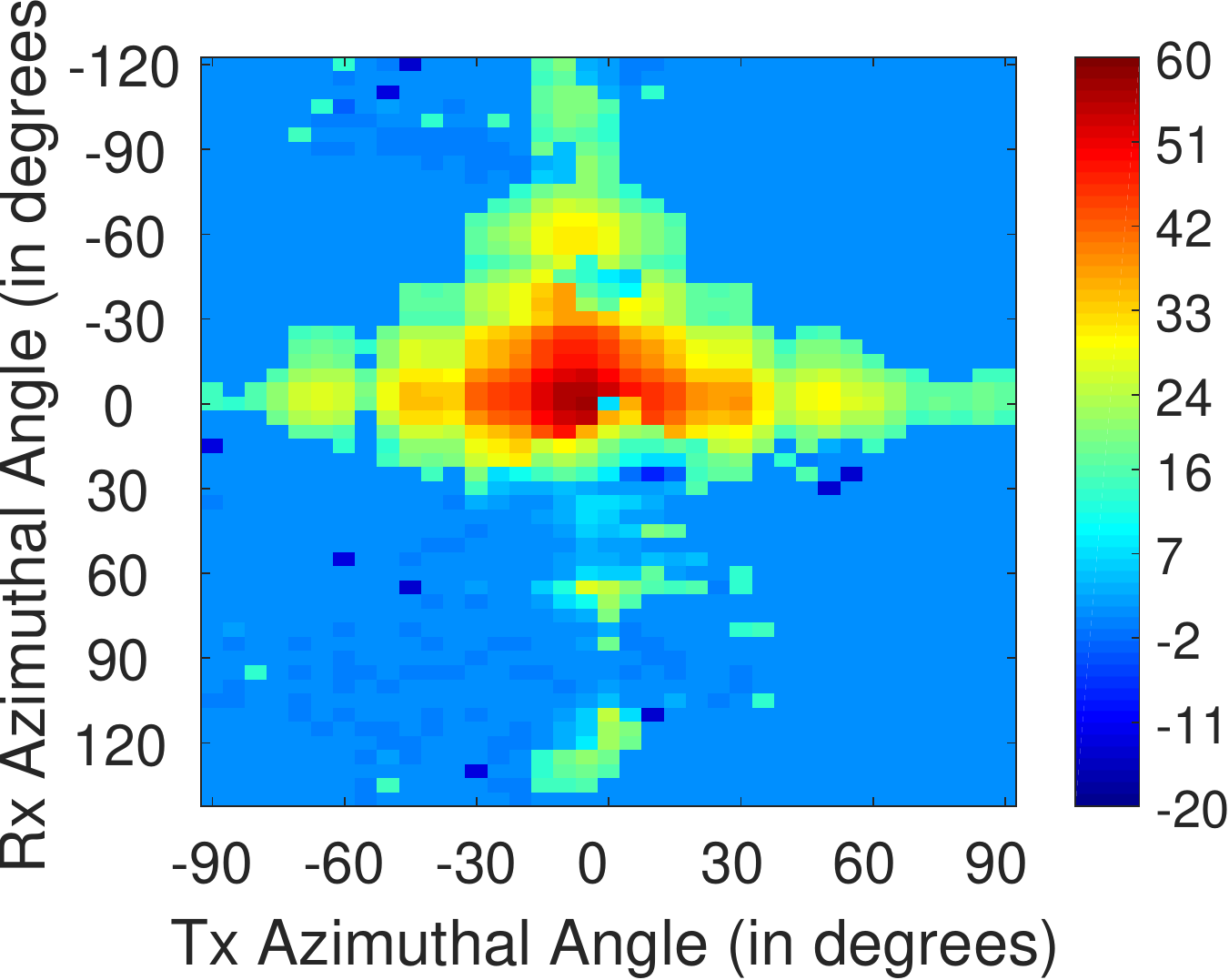}}
        \\
        \subfloat[No reflector (no tripod).]{\includegraphics[width=.2\textwidth,height=.2\textwidth]{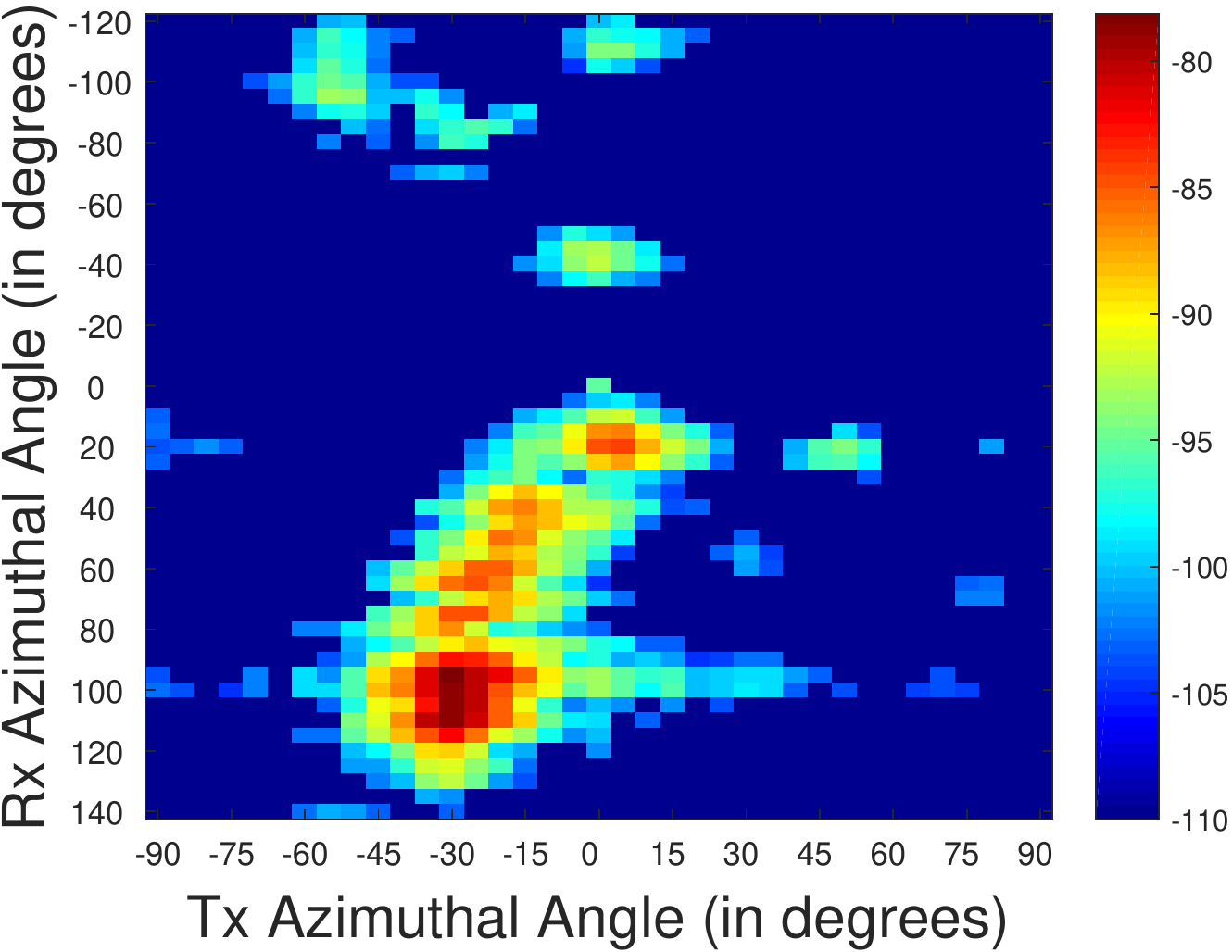}}
       \subfloat[Metal reflector.]{\includegraphics[width=.2\textwidth,height=.2\textwidth]{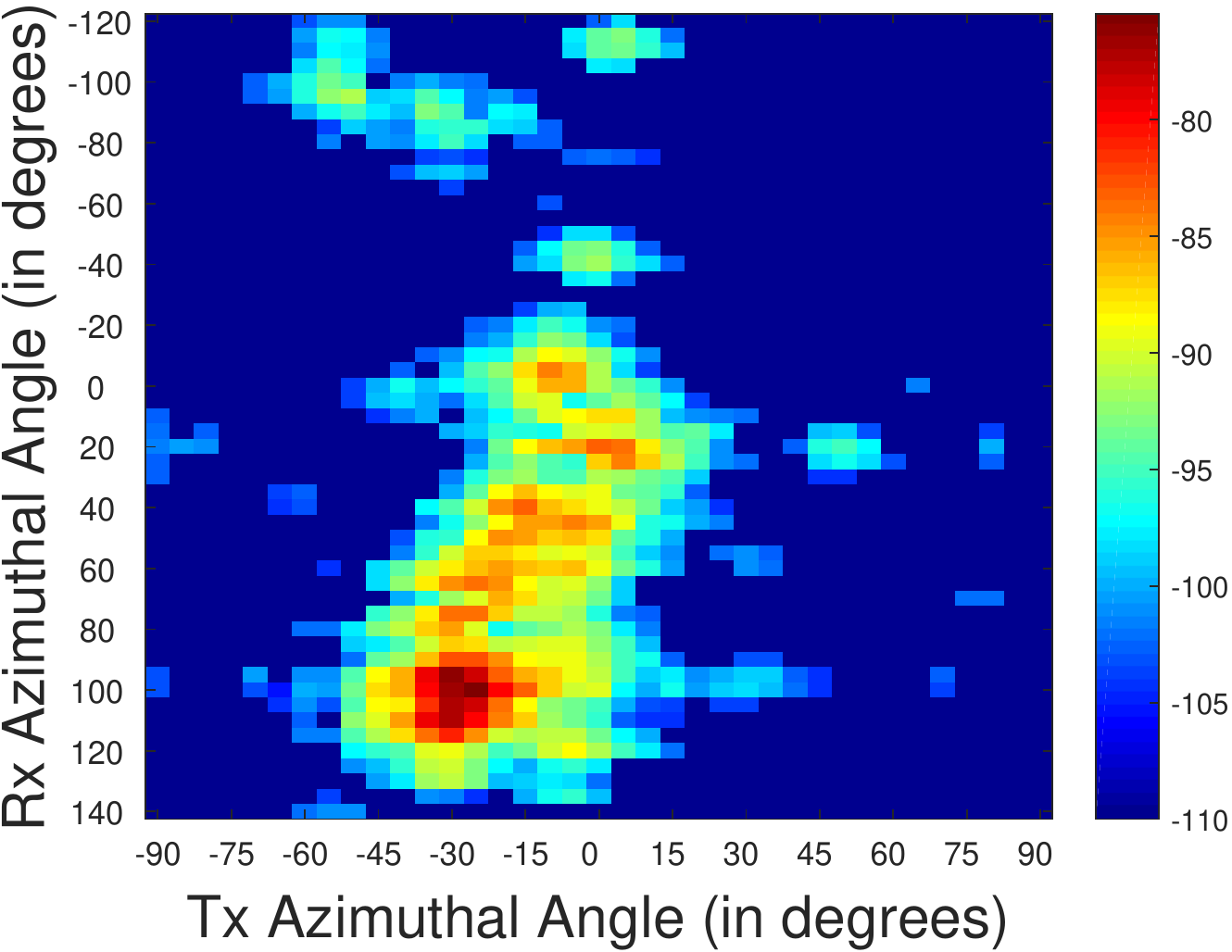}}
         \subfloat[Transparent reflector.]{\includegraphics[width=.2\textwidth,height=.2\textwidth]{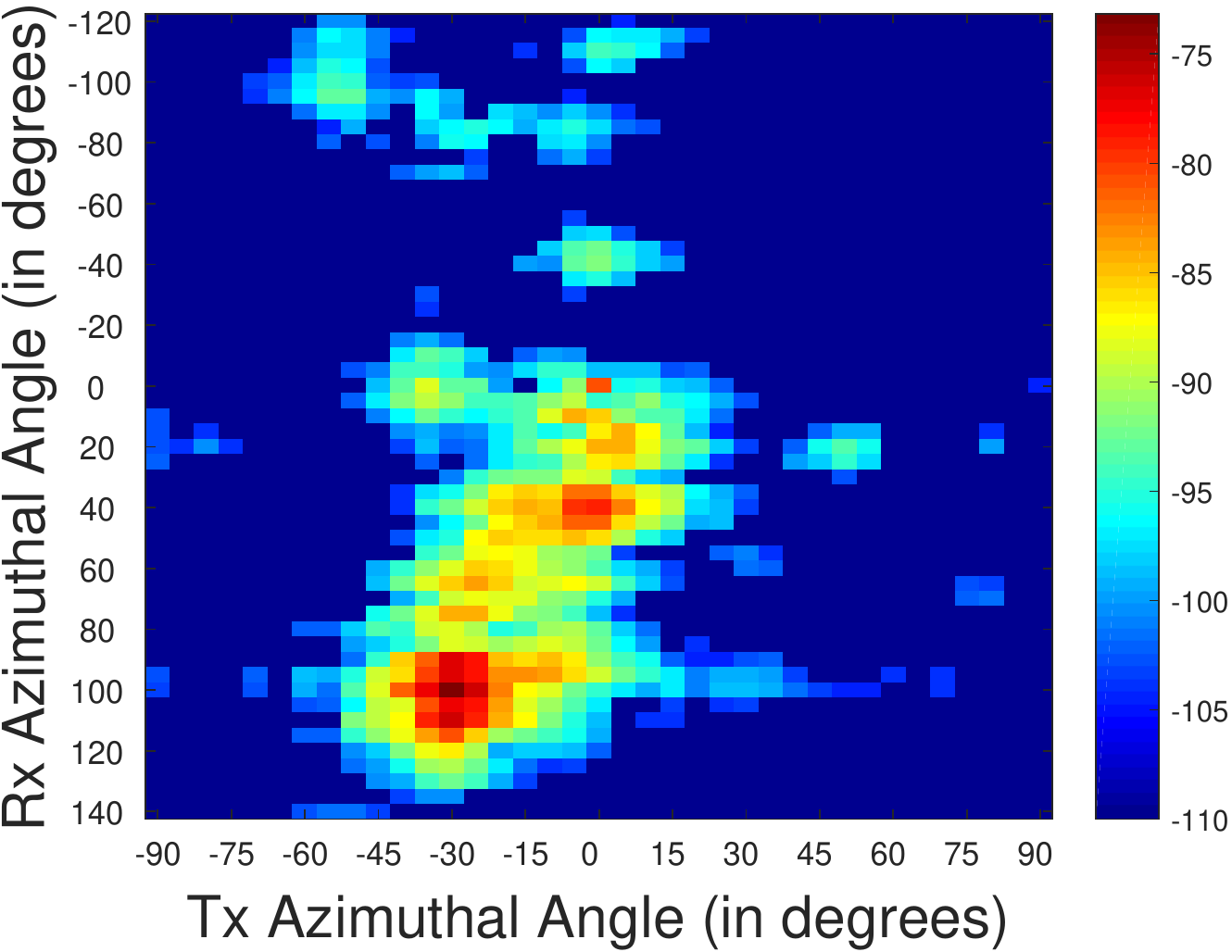}}
        \subfloat[MRC: $P_{\max}$ = 25.55 dB.]{\includegraphics[width=.2\textwidth,height=.2\textwidth]{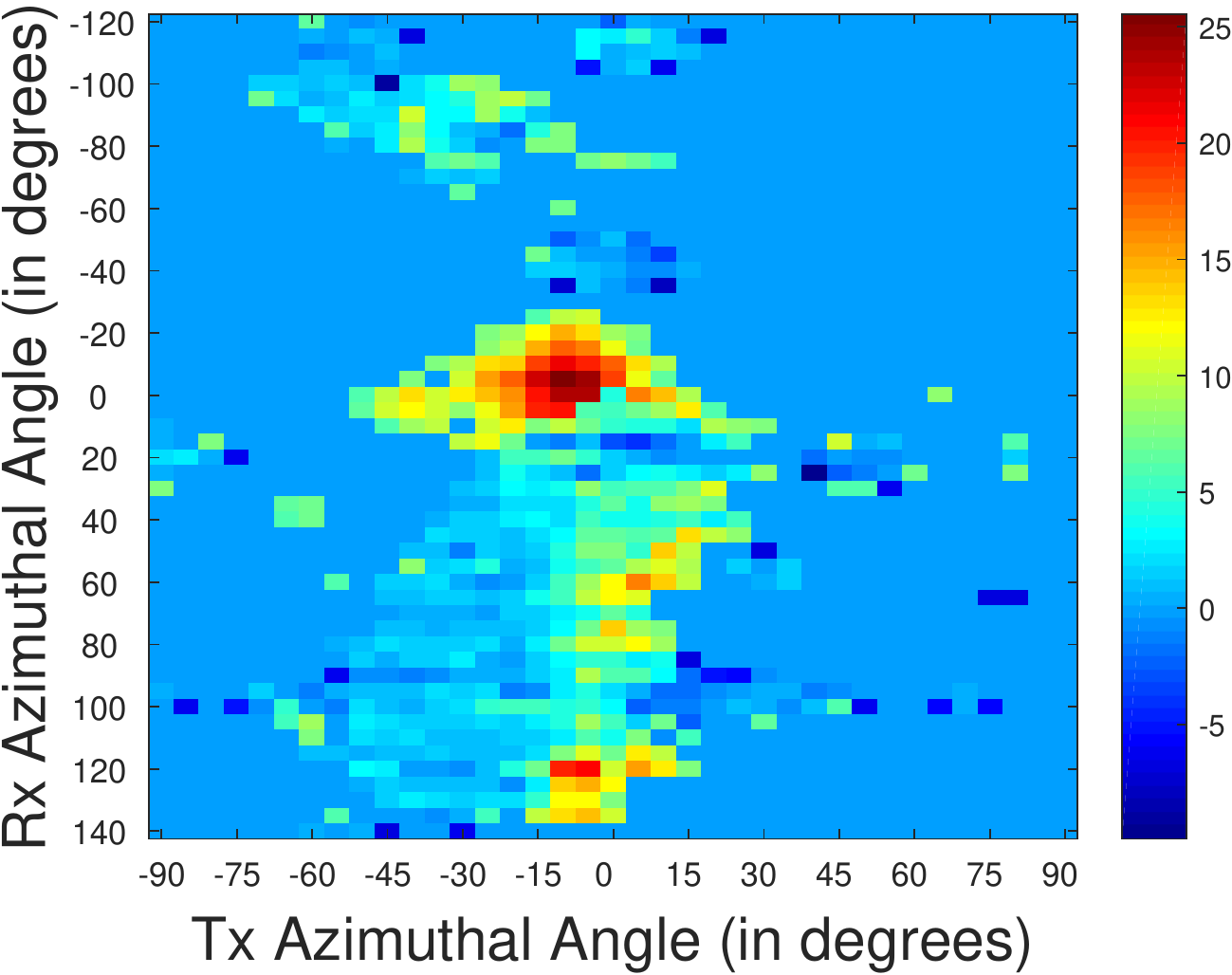}}
        \subfloat[TRC: $P_{\max}$ = 23.24 dB.]{\includegraphics[width=.2\textwidth,height=.2\textwidth]{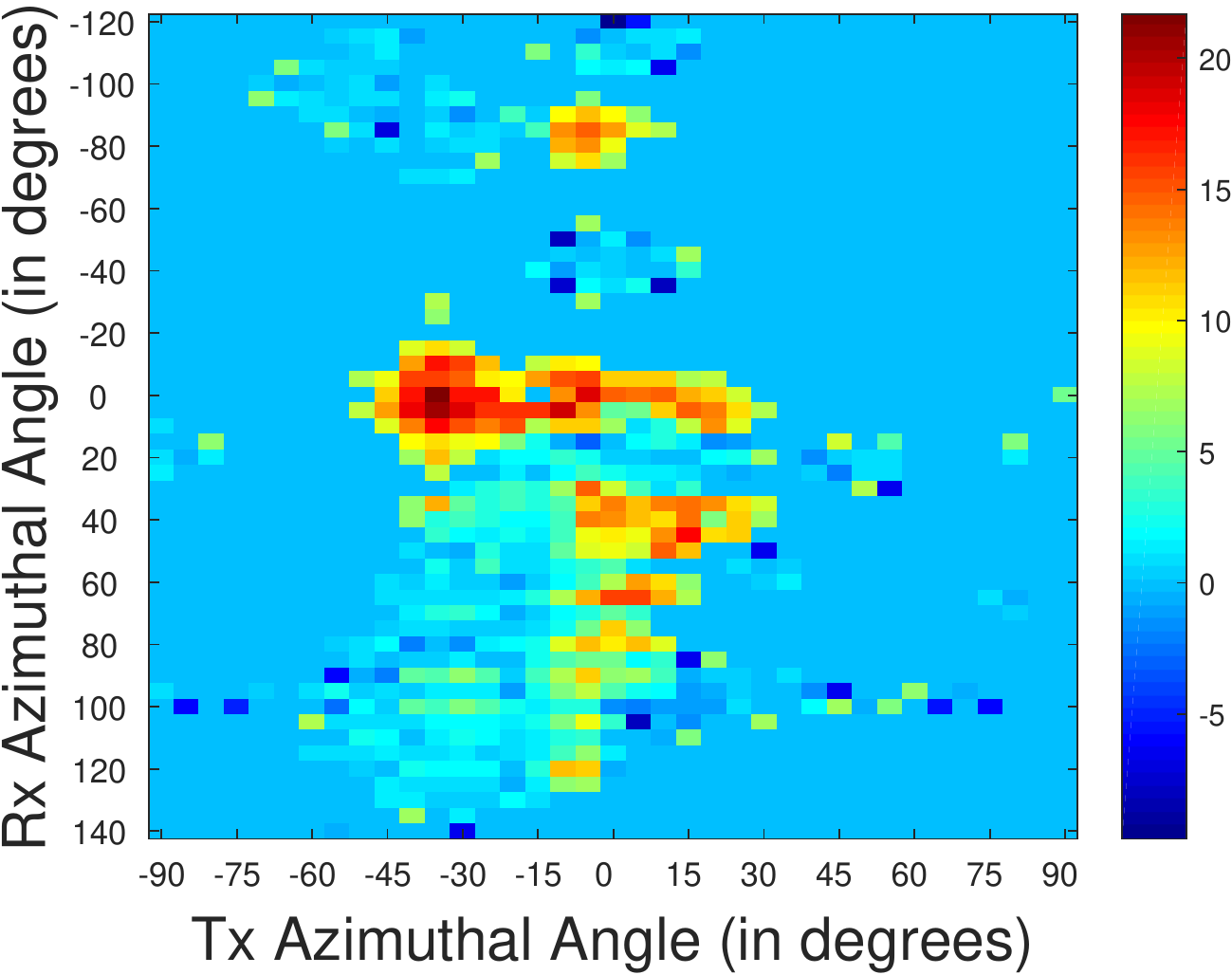}}
    \caption{28 GHz: first row (a)-(e), 39 GHz: second row (f)-(j), 120 GHz: third row (k)-(o), 144 GHz: fourth row (p)-(t). The colormap indicates the received signal power with units in dBm. For 120 GHz and 144 GHz, the colormap {indicates relative power as the power levels are not calibrated.} }
    \label{fig:Heatmap_All_freq}
\end{figure*}

\section{Experimental Results}\label{Sec:Results}
In this section, we present the experimental results and draw critical insights into the propagation characteristics of the transparent reflector. We do so by comparing it against the performance of commonly found indoor materials. In general, we treat the metal reflector as a benchmark due to its ideal propagation characteristics. For the measurements, the transmit power $P_\text{Tx}$ was set to -10 dBm for all the experiments unless mentioned otherwise, and {the bandwidth of the transmitted signal was set to be 1.5 GHz.}  All the results obtained in this section are averaged over the multiple snapshots of each~measurement.


\subsection{Penetration Loss Results}\label{Sec:Results_PL}
Fig.~\ref{fig:Penetration_Loss} illustrates the averaged $L_\text{Pen.}$ results of different materials and is obtained as discussed in Section~\ref{SubSec:Penetration}. It should be emphasized that Fig.~\ref{fig:Penetration_Loss}(a) shows the absolute $L_\text{Pen.}$, whereas, Fig.~\ref{fig:Penetration_Loss}(b) denote the normalized $L_\text{Pen.}$ per centimeter thickness of the material. As seen in Fig.~\ref{fig:Penetration_Loss}(a), the ceiling tile offers the lowest attenuation at all four frequencies, followed by other materials (such as clear glass, drywall, and plywood) with different attenuation factors not exceeding 7.5 dB at all four frequencies. Amongst all the materials, metal has the highest $L_\text{Pen.}$ followed by the transparent reflector. The high $L_\text{Pen.}$ of metal is obvious because of its small penetration depth and skin depth. At 28 GHz and 39 GHz, metal reflector resulted in a {weak obstructed LoS}. However, at 120~GHz and~144~GHz, there was no obstructed LoS (no MPC) through the metal, even at the maximum power of 10 dBm supported by the sub-THz equipment. Thus, we have capped the $L_\text{Pen.}$ as~70 dB (maximum dynamic range supported by the system) in Fig.~\ref{fig:Penetration_Loss}(a), which can be treated as a lower bound. 

In general, the trend is intuitive: higher the frequency higher the attenuation of the material, excluding metal and transparent reflector. Fig.~\ref{fig:Penetration_Loss}(b) reveals a similar trend as demonstrated in~\cite{kairui}. 
An interesting and important result is the $L_\text{Pen.}$ of the transparent reflector. Despite being transparent, it offers higher penetration loss than common indoor materials except for metal, with an obvious cosmetic advantage. For example, the ceiling tile alone in an indoor setup would leak the power if the Tx-beam is pointed upward towards the ceiling tile. However, covering the ceiling tile with transparent reflectors would not leak the power due to high penetration loss, but will rather preserve the radio waves within the room resulting in a better radio signal strength.

\subsection{Reflection Results}\label{Sec:Results_Reflection}
Fig.~\ref{fig:Heatmap_All_freq} illustrates the heatmap of the total power received at the Rx for each of the Tx/Rx rotating gimbal angle bin. The figure contains four rows, where each row contains the results for each frequency, in the order of 28 GHz, 39 GHz, 120 GHz, and 144 GHz, respectively. Further, each row contains five columns, where the first column corresponds to no reflector (no tripod) scenario, followed by the heatmaps of received power with the metal and transparent reflector, respectively. The last two columns denote the contribution from the metal and the transparent reflector contribution (MRC/TRC) alone, respectively. These two plots help to compare the performance of the reflectors and they are acquired by subtracting the results for the metal/transparent reflector from the results with no reflector,~respectively.

A pragmatic assessment of the first column across all the four frequencies reveals that there is a strong obstructed LoS path penetrating through the wall peaking at a Tx/Rx angle of (-30$^\circ$, 90$^\circ$); see Fig.~\ref{fig:Exp_Setup}(b)-(c) for the obstructed wall and the measurement setup. Further, as the operating frequency increases, the number of Tx/Rx angle combinations for possible NLoS links decreases due to the increased path loss. We emphasize that the absolute power levels across different frequencies depend on the varying capabilities at those frequencies, ranging from different antenna gains and beamwidths, different beam spillage ratio, among others (see Section~\ref{Sec:Equipment_Details}).

Heatmaps in Columns 2 and 3 reveal that having a reflector in the evaluated scenario provides an NLoS link with a similar performance as the obstructed link at all four frequencies (power peaking at a Tx/Rx angle of (0$^\circ$, 0$^\circ$) ). This signifies the promising benefits of passive reflectors at mmWave and sub-THz bands. Amongst the metal and transparent reflector, at 28 GHz and 120 GHz, TRC contributes in a slightly higher power than MRC, and vice-versa at 39 GHz and~144~GHz. The inconsistency in result may be due to the manual replacement of the reflectors during measurements that may have caused a slight change in the orientation. 
Nevertheless, the maximum power deviation does not exceed~4~dB. Overall, these preliminary but encouraging results suggest that the transparent reflectors perform similar to the metal reflector. Our other indoor experimental results (not presented in this~paper) are in good agreement with our presented results and have shown~a~similar~trend. 


\section{Conclusion}\label{Sec:Conclusion}
{In this work, we demonstrated for the first time in the literature that transparent reflectors can be used as passive reflectors with similar performance as metallic reflectors for enhancing the wireless coverage at mmWave and sub-THz bands.} In particular, we performed measurements at 28 GHz, 39 GHz, 120 GHz, and 144 GHz, characterized penetration loss of common indoor materials, and compared reflection characteristics of the transparent reflector against the metal reflector. The measurements results suggest that the transparent reflector, apart from an obvious advantage of transparency, has higher penetration loss than the indoor materials such as ceiling tile, clear glass, drywall, and plywood,  and aids in preserving the radio waves within the indoor environment. Also, it performs similarly to metal reflectors in terms of reflection. {Our future work includes experiments to characterize the various radio properties of the transparent reflectors at different geometrical positions.}

\vspace{-.15cm}
\bibliographystyle{IEEEtran}
\bibliography{IEEEabrv,references}

\end{document}